\documentclass[numberedappendix]{emulateapj}

\shorttitle{New reddening maps of the MCs}
\shortauthors{Haschke, Grebel \& Duffau}

\begin{document}
   \title{New optical reddening maps\\of the Large and Small Magellanic Clouds}


\author{Raoul Haschke
          \&
	  Eva K. Grebel
	  and
          Sonia Duffau
	   }

\email{haschke@ari.uni-heidelberg.de}

\affil{Astronomisches Rechen-Institut, Zentrum f\"ur Astronomie der Universit\"at Heidelberg,
	      M\"onchhofstrasse 12-14, D-69120 Heidelberg, Germany}\

\begin{abstract}
We present new reddening maps of the Small and Large Magellanic Cloud based on the data of the third phase of the Optical Gravitational Lensing Experiment (OGLE\,III). We have used two different methods to derive optical reddening maps. We adopt a theoretical mean unreddened colour for the red clump in the SMC and LMC, respectively. We subdivide the photometric data for both Clouds into subfields and calculate the difference between the observed red clump position and the theoretical value for each field, which provides us with the reddening value in $(V-I)$. Furthermore reddening values are obtained for 13490 LMC RR\,Lyrae\,ab and 1529 SMC RR\,Lyrae\,ab stars covering the whole OGLE\,III region of the MCs. The observed colours $(V-I)$ of the RR\,Lyrae stars are compared with the colour from the absolute magnitudes. The absolute magnitude of each RR\,Lyrae star is computed using its period and metallicity derived from Fourier decomposition of its lightcurve.\\\noindent\hspace*{1em}
In general we find a low and uniform reddening distribution in both Magellanic Clouds. The red clump method indicates a mean reddening of the LMC of $E(V-I) = 0.09 \pm 0.07$\,mag, while for the SMC $E(V-I) = 0.04 \pm  0.06$\,mag is obtained. With RR\,Lyrae stars a median value of $E(V-I) = 0.11 \pm 0.06$\,mag for the LMC and $E(V-I) = 0.07 \pm 0.06$\,mag for the SMC is found. The LMC shows very low reddening in the bar region, whereas the reddening in the star-forming leading edge and 30\,Doradus is considerably higher. In the SMC three pronounced regions with higher reddening are visible. Two are located along the bar, while the highest reddening is found in the star-forming wing of the SMC. In general the regions with higher reddening are in good spatial agreement with infrared reddening maps as well as with reddening estimations of other studies. The position-dependent reddening values from the red clump method are available via the German Astrophysical Virtual Observatory interface at \url{http://dc.zah.uni-heidelberg.de/mcx}
\end{abstract}


%

\section{Introduction}
\label{introduction}
%
\defcitealias{Dobashi08}{D08} 
\defcitealias{Dobashi09}{D09} 
\defcitealias{Zaritsky02}{Z02} 
\defcitealias{Zaritsky04}{Z04}
\defcitealias{Subramaniam05}{S05} 
\defcitealias{Subramanian09}{S09} 
\defcitealias{Pejcha09}{PS09}

The gas rich and star-forming Small Magellanic Cloud (SMC) and Large Magellanic Cloud (LMC) are the most massive galaxies orbiting the Milky Way. They are used in many ways as benchmarks, for instance, for studies of star formation at lower metallicity \citep[e.g.][]{Cignoni09, Glatt08a, Sabbi09} and as calibrators for the cosmological distance scale \citep{Freedman01}. Large optical and infrared imaging surveys of the Magellanic Clouds (MCs), such as the Optical Gravitational Lensing Experiment \citep[OGLE,][]{Udalski92, Udalski08b, Udalski08a}, the Magellanic Clouds Photometric Survey \citep[MCPS,][]{Zaritsky97}, the Massive Compact Halo Objects (MACHO) survey \citep[e.g.][]{Alcock00, Cook92}, the Exp{\'e}rience pour la Recherche d'Objets Sombres \citep[EROS,][]{Aubourg93} or the Two Micron All Sky Survey \citep[2MASS,][]{Skrutskie06}, have been completed in recent years. These surveys provide invaluable information about the stellar content of the MCs . But for interpreting the data of these surveys, for deriving star formation histories and for stellar population studies in general the knowledge of the reddening is crucial.\\\noindent\hspace*{1em}
Reddening is caused by gas and dust located between the emitting object and the observer. This leads to dimmer observed magnitudes and redder colours than emitted by the object. The magnitude of absorption and scattering depends on the grain sizes of the dust and on the wavelength of the passing light. For different wavelengths the absorption by dust has a differing effectiveness. The resulting change of the observed colour of the star is defined as $E(\lambda_1-\lambda_2) = (\lambda_1-\lambda_2)_{\mathrm{obs}} - (\lambda_1-\lambda_2)_{\mathrm{intrinsic}}$ and is quoted in the literature as colour excess, selective extinction or reddening. The total extinction $A_{\lambda}$ can be calculated by adopting an extinction law. Using stars from the local neighbourhood \citet{Cardelli89} found that an extinction law can be introduced that depends on just one parameter: $R_V = A_V/E(B-V)$, which has a typical value of $R_V = 3.1$. \citet{Gordon03} tested various lines of sight to test the local extinction laws for more distant fields in the Galaxy, in the LMC and the SMC. They found that for some fields the law is indistinguishable from the result by \citet{Cardelli89}, but many fields show considerable differences from the local extinction law. This is explained by the different environments probed by the authors. However, \citet[their Figure\,10]{Gordon03} show that for the visual wavelength range, which is of interest in our current study, the differences in the extinction laws are quite small. For the visible wavelength range \citet{Zagury07} confirms that the MCs have a similar extinction law as the Galaxy, while he suggests that for regions with deviations other explanations than a differing reddening law might also be valid. Even though the dust content of the Galaxy is far from being understood these studies show that at optical wavelength adopting the extinction law by \citet{Cardelli89} is generally a valid assumption. \\\noindent\hspace*{1em}
Reddening maps of the MCs have been obtained by several groups. With infrared data from the 2MASS \citet[herafter D08 and D09]{Dobashi08, Dobashi09} explored the dust content of the LMC and SMC, which can be translated into and compared with reddening maps. The optical data of the MCPS were analysed fitting photometric model predictions of stellar luminosity, effective temperatures and extinctions to apparent U, B, V and I magnitudes for the stars in the MCs. In \citet{Zaritsky99} an extinction map is constructed by finding the best match between the theoretical values and the observations for a fraction of the LMC field. The whole dataset of the MCPS results in wide area reddening maps of the MCs that are presented in \citet[hereafter Z02 for the SMC and Z04 for the LMC]{Zaritsky02, Zaritsky04}. \\\noindent\hspace*{1em}
%
%
For OGLE\,II (see Section\,\ref{data}) reddening maps from estimates based on red clump (RC) stars \citep{Udalski99a, Udalski99b} were provided. A similar technique was adopted by \citet[hereafter S05]{Subramaniam05} and \citet[hereafter S09]{Subramanian09}. They measured the position and apparent shift of the RC in the colour magnitude diagram (CMD) with respect to the theoretically unreddened values to analyse the spatial distribution of the reddening in the V- and I-band of the LMC covered by the OGLE\,II and the OGLE\,III surveys.  \\\noindent\hspace*{1em}
\citet{Sturch66} introduced a completely different approach of measuring reddening values. He proposed to use the difference between the observed colour and the intrinsic colour of RR\,Lyrae stars as an indicator for the reddening towards these stars. RR\,Lyraes are standard candles with distinct relations to calculate absolute magnitudes from the observable parameters apparent magnitude, period and metallicity. These can be used to obtain the intrinsic colour. \citet[herafter PS09]{Pejcha09} used this technique for RR\,Lyrae stars observed by OGLE\,III to derive a reddening map of the LMC. For their computations they assume that that all RR\,Lyrae stars share a common metallicity. \\\noindent\hspace*{1em}
Here we present new optical reddening maps for the MCs derived via two different methods. In Section\,\ref{data} we present the data. In Section\,\ref{red_clump} we describe the RC method and the resulting reddening maps of the MCs. We use the method of measuring the red clump, introduced by \citet{Wozniak96}, in small subfields of the MCs to determine new reddening estimates for both Clouds. As mentioned earlier, in OGLE\,II the reddening was derived using the RC method (\citetalias{Subramaniam05} and \citet{Udalski99a}), while OGLE\,III leaves this task to the user. Since \citetalias{Subramanian09} sample the OGLE\,III area with a grid of constant subfield size and then apply the RC method in each subfield, some fields are not sufficiently populated to yield a reddening value. The individual reddening values of \citetalias{Subramanian09} are not available in tabular form. Using an adjusting field size we are able to circumvent the undersampling problem. \\\noindent\hspace*{1em}
In Section\,\ref{RR_Lyrae} we present the method and results for the RR\,Lyrae reddening. We calculate the absolute magnitudes of the RR\,Lyrae taking their metallicity into account, which we obtain from Fourier decomposition of their light curves. \\\noindent\hspace*{1em}
The discussion of the reliability of our reddening determinations and the comparison to other reddening maps are presented in Section\,\ref{discussion}. Section\,\ref{summary} summarises all results.

%

\section{Data}
\label{data}
The OGLE collaboration has been taking data since 1992. In the first phase of OGLE the 1m Swope telescope at Las Campanas Observatory, Chile, was used. With the start of phase\,II in 1995 OGLE's own 1.3\,m telescope at Las Campanas became operational. In the first two phases OGLE used a $2048 \times 2048$\,pixel camera with a field of view of $15' \times 15'$. The data were taken preferentially in the I-band. In total 2.4 square degrees in the SMC and 4.5 square degrees in the LMC \citep{Udalski92, Udalski97} were covered. In 2001 the third phase (OGLE\,III) began and ran until mid 2009. With a mosaic of 8 CCDs with a total of $8192 \times 8192$\,pixels a field of view of $35' \times 35'$ was observed at once \citep{Udalski03}. Overall the fields cover 14 square degrees for the SMC and nearly 40 square degrees for the LMC. These observations provide photometry and astrometry in the I- and V-band for 6.2\,million stars in the SMC and for about 35\,million stars for the LMC \citep{Udalski08a, Udalski08b}. The point source catalogues\footnote{The catalogues are available and downloaded from \url{http://ogle.astrouw.edu.pl/}} are divided into 40 fields for the SMC and 116 fields for the LMC. The 41st SMC field targets the massive Galactic globular cluster 47\,Tuc and is ignored in our study.

%

\section{Reddening based on the red clump}
\label{red_clump}
\subsection{Method}
\begin{figure}
\centering 
 \includegraphics[width=0.50\textwidth]{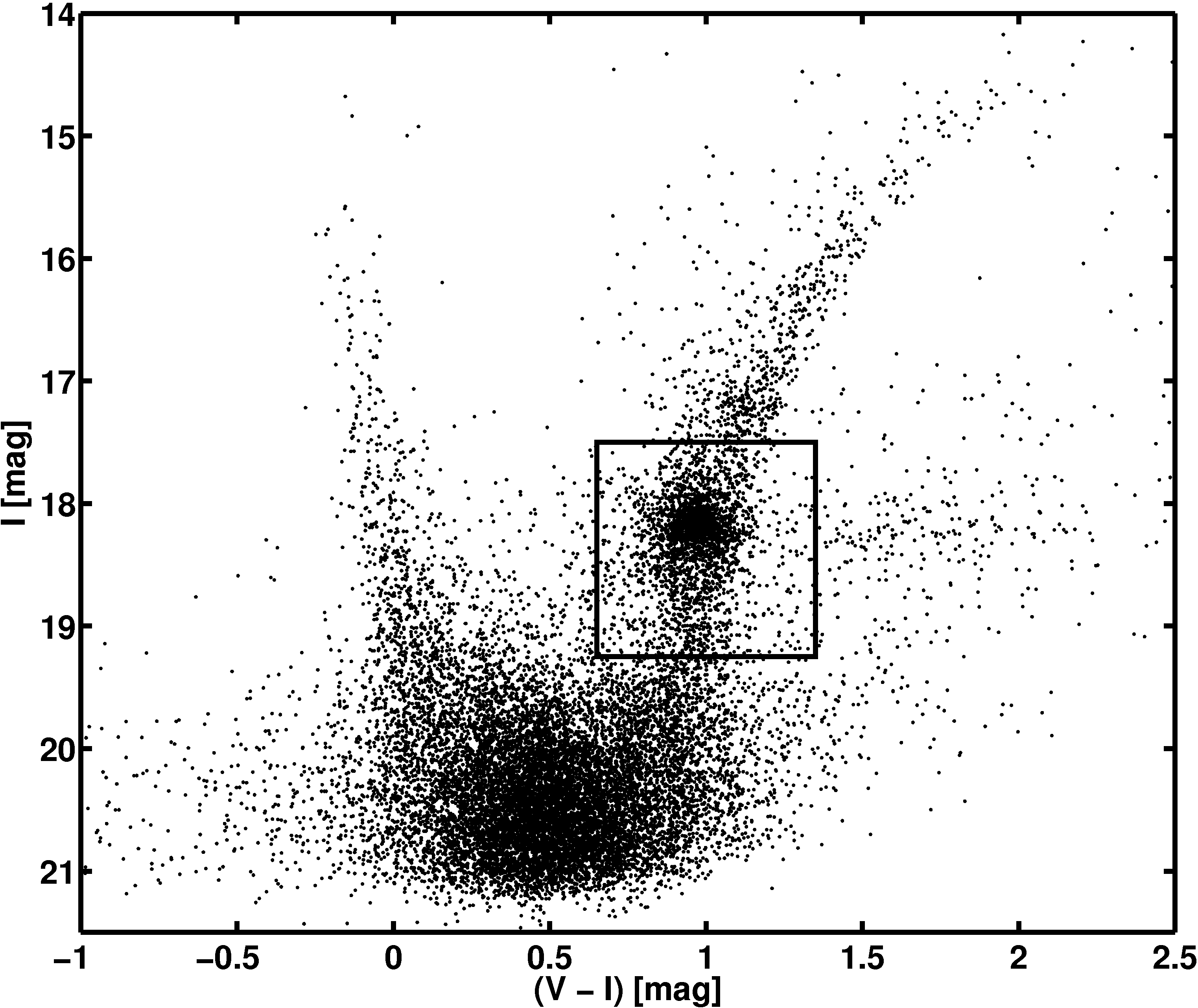}
 \caption{A typical CMD of a LMC subfield. This field is located at $\alpha = 81^\circ32'$ and $\delta = -69^\circ33'$ with a projected size of $9' \times 9'$. It contains 18268 stars. 2661 stars are in our red clump selection box, represented by the continuous line.} 
 \label{red_clump_in_CMD} 
\end{figure}

\begin{figure}
\centering 
 \includegraphics[width=0.50\textwidth]{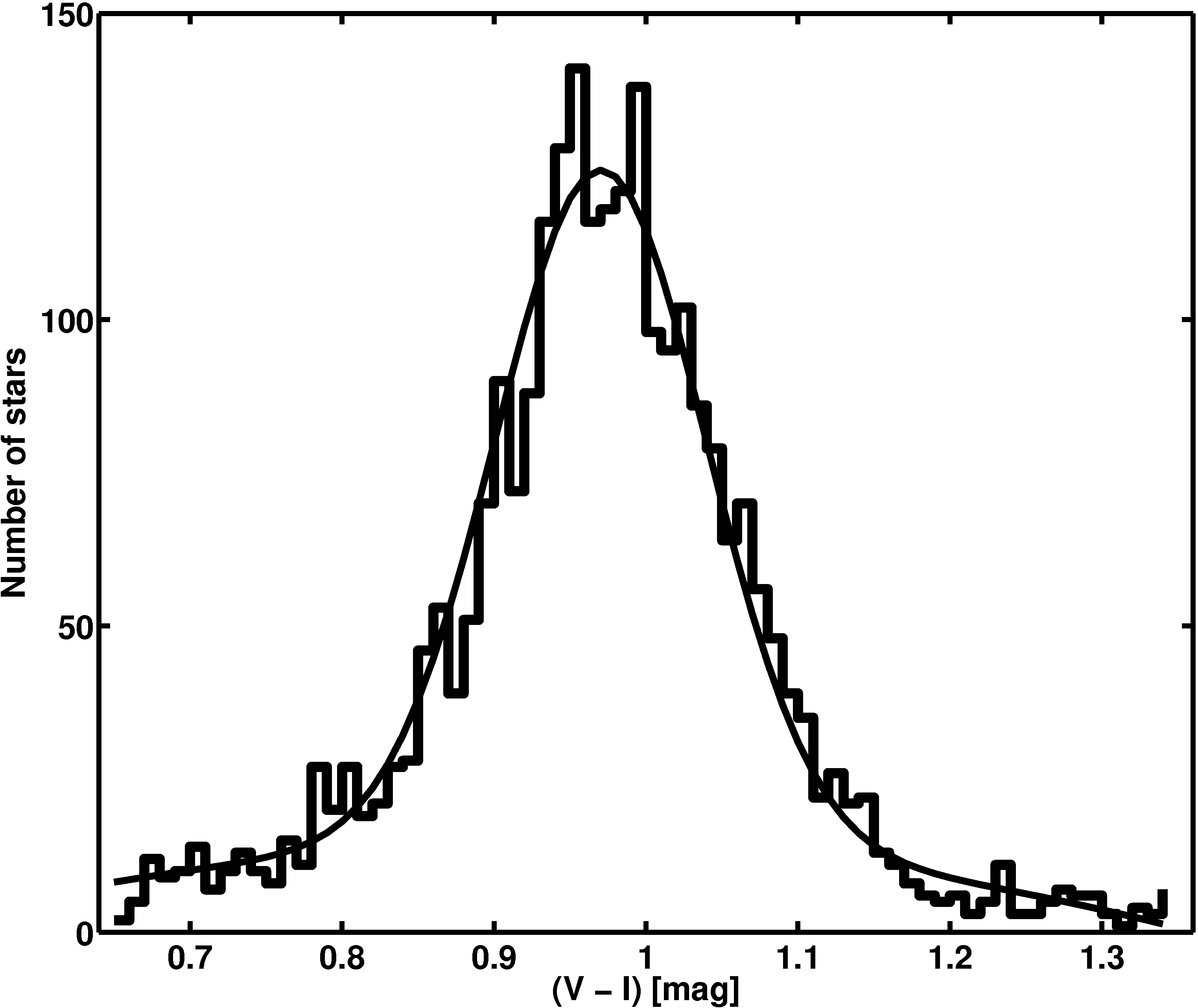}
 \caption{Histogram distribution of the RC stars in one of the examined fields, which is centred on $\alpha = 81^\circ32'$ and $\delta = -69^\circ33'$., The solid line represents the fit to the distribution and the $\chi^2$ of the fit in this field is 1.06. We find a mean colour of $(V-I) = 0.97$\,mag for the 2661 stars.} 
 \label{histogram_mit_fit} 
\end{figure}

Our goal is to infer a reddening map for both Clouds using the mean position of the RC in a CMD. \\\noindent\hspace*{1em}
For this purpose we define a selection box in colour-magnitude space. All stars between $17.50 \leq I \leq 19.25$\,mag and $0.65 \leq (V-I) \leq 1.35$\,mag are taken into account for the determination of the mean locus of the RC in the LMC fields (see Figure\,\ref{red_clump_in_CMD}). These boundaries correspond to the selection box used by \citetalias{Subramaniam05} (\citetalias{Subramanian09} adopts the methods and specifications of \citetalias{Subramaniam05}). For the SMC the limits in colour are chosen as in the LMC, but due to the greater distance, the box is shifted by 0.3\,mag towards fainter I-band magnitudes. \\\noindent\hspace*{1em}
To find an adequate number of stars the size of each field has to be carefully chosen. With a low number of stars the uncertainty in determining the mean colour would increase, while too large a number of stars prevents a map with details on smaller scales. We want to define a grid that allows us to obtain reddening information on scales as small as possible across the OGLE\,III field. Due to the density differences between the central parts and the outskirts of the Clouds we have to adjust the subfield sizes in dependence of the absolute number of stars in a given area. Our angular resolution is therefore variable and only driven by the available number of stars as listed in Table\,\ref{subfieldsizes_table}. \\\noindent\hspace*{1em}
The individual fields of the SMC and LMC defined by the OGLE collaboration are large and contain many more stars than needed to find a significant red clump in the colour-magnitude diagram. Therefore our first step for both galaxies is to divide each OGLE field in three, nearly quadratic, subfields. Due to the very different numbers of stars in these subfields we make further subdivisions dependent on the stellar density. If the total number of stars in the selection box in a subfield exceeds a certain threshold, this field is divided further, taking care that the number of RC stars per field does not drop below a few hundred stars. If a quadratic subfield of $36 \times 36\, \mathrm{arcmin}^2$ contains for example more than 40,000 RC stars this field is cut into 64 equally sized fields with an average size of $4.5 \times 4.5\,\mathrm{arcmin}^2$. For different stellar densities Table\,\ref{subfieldsizes_table} lists the number of stars of the quadratic subfield in the first column, while the number of equally sized subfields and the dimensions of the finally evaluated fields are given in the second and third column, respectively. \\\noindent\hspace*{1em}
\begin{table}
\begin{center}
\caption{Sizes of the finally examined subfields in dependence of the number of stars in the evaluated OGLE\,III field.\label{subfieldsizes_table}}
\begin{tabular}{ccc}
\tableline\tableline
$\#$ of Stars & $\#$ of subfields & size [arcmin] \\    
\tableline                       
$>$ 40,000 & 64 & 4.5 $\times$ 4.5 \\ 
$>$ 12,000 & 16 & 9 $\times$ 9\\ 
$\geqq$ 3,000 & 4 & 18 $\times$ 18 \\ 
$<$ 3,000 & 1 & 36 $\times$ 36 \\ 
\tableline                                  
\end{tabular}
\end{center}
\end{table}
In total we divide the LMC into 3174 subfields, while the SMC reddening is examined in 693 subfields. The number of RC stars in each examined subfield varies in the LMC from 260 to 4046, with an average number of 1257 RC stars in each field. For the SMC the least populated field has 477 RC stars, while the population reaches its maximum at 3635. On average 1318 RC stars are located in each field. \\\noindent\hspace*{1em}
In theory all RC stars have a well-defined mean colour $(V-I)_0$. This colour only depends on the mean metallicity and the age of the population. \citet{Girardi01} examined theoretical RC colours in dependence of metallicity and computed mean values for $(V-I)_0$. For the LMC we adopt $(V-I)_0 = 0.92$\,mag ($z = 0.004$) from the work by \citet{Olsen02}, who inspected red clump stars to test for a warp in the LMC. For the SMC, due to the lower metallicity \citep[$z \sim 0.0025$;][]{Cole98, Glatt08b}, $(V-I)_0 = 0.89$\,mag is used. In \citet{Udalski98} a similar value for the unreddened RC of the SMC was used.\\\noindent\hspace*{1em}
Due to reddening effects by interstellar dust and gas the mean colour of the clump is shifted redwards, which we denote as $(V-I)_{\mathrm{obs}}$. By calculating the difference between the theoretical value and the observed colour of the field the reddening can be obtained with $E(V-I) = (V-I)_{\mathrm{obs}} - (V-I)_0$.  \\\noindent\hspace*{1em}
%
%
To find the mean colour of the RC the histogram of the RC colour distribution (Figure \ref{histogram_mit_fit}) is fitted with a Gaussian plus a second order polynomial. The maximum of this function is assumed to be the mean colour of the clump. Different effects contribute to a broadening of the peak, such as the intrinsic scatter around the mean locus, photometric errors, the range of ages, differential reddening, the presence of binaries or differences in metallicity as well as contributions of red giants and red horizontal branch stars. \\\noindent\hspace*{1em}
The mean reduced $\chi^2_{\mathrm{mean}}$ of the fits is 1.03 for the LMC, indicating that a Gaussian provides a good approximation to the data. Fields with a reduced $\chi^2 \geq 3$ were removed from the sample and then inspected by eye to check whether the mean location of the RC was correctly identified. Only 4 of the 3174 subfields in the LMC needed further inspection. For the SMC 34 out of 693 fields had a reduced $\chi^2 \geq 3$, while the $\chi^2_{\mathrm{mean}}$ is 1.52. The deviant LMC and SMC subfields are mostly affected by a strong red giant branch, which leads to a second bump in the histogram. The small displacements of the mean colour are taken into consideration in the calculation of the uncertainties and hence no fields were rejected.  \\\noindent\hspace*{1em}
\newpage
\subsection{Results for the LMC}
\begin{figure}
 \centering 
 \includegraphics[width=0.50\textwidth]{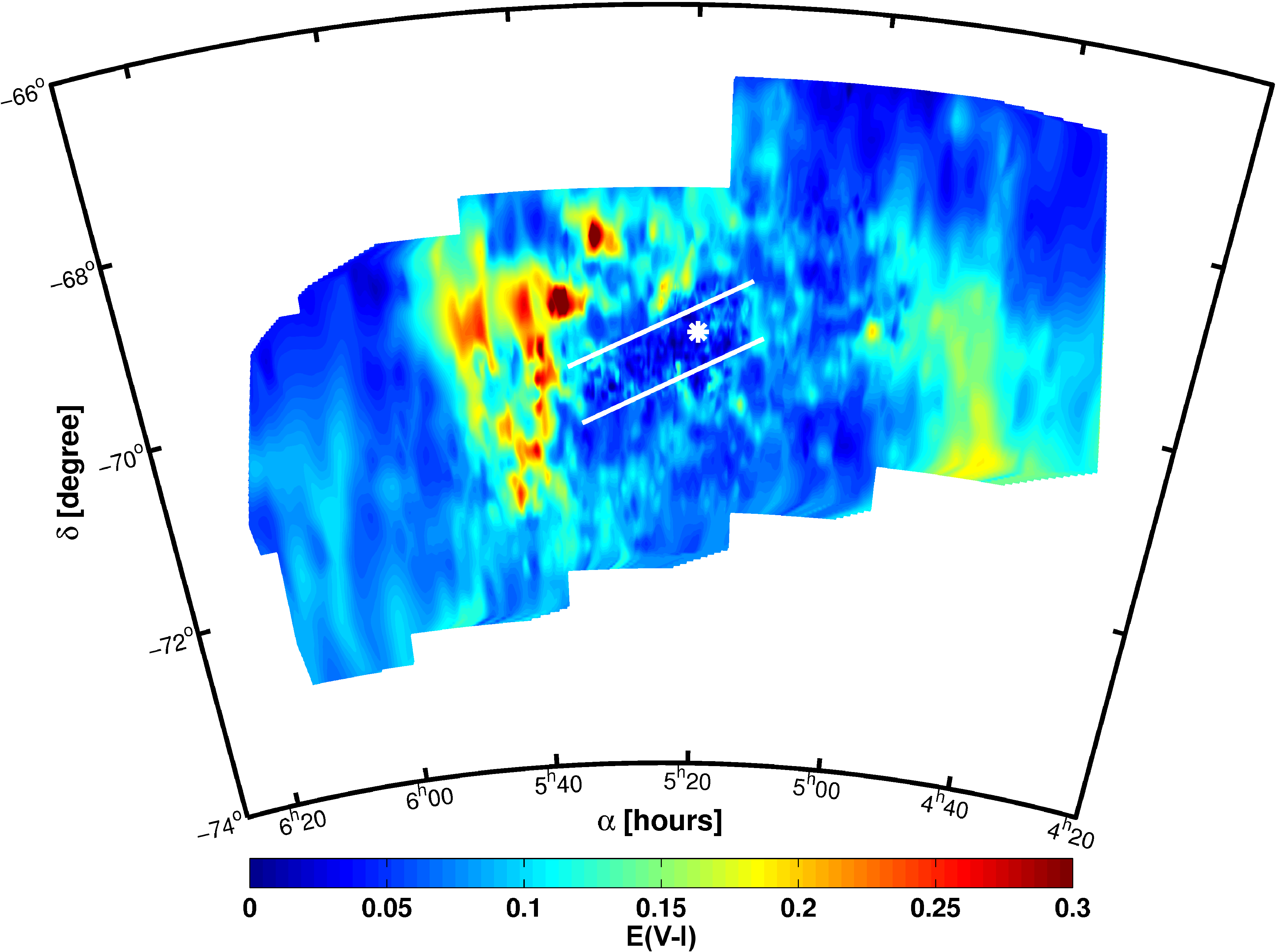}
 \caption{LMC reddening distribution. This contour map shows high reddening values in red and low reddening in blue. In the bar region of the LMC, marked with  two diagonal white lines, the reddening is very low. Only along the leading edge of the LMC we find comparatively high reddening with values up to $E(V-I) = 0.43$\,mag. The white star represents the optical centre of the LMC ($\alpha = 5^{h}19^{m}38^{s}$ and $\delta = -69\degr27'5\farcs2$) found by \citet{Vaucouleurs72}. Reddening values in 30\,Doradus and the leading edge are only lower limits due to high differential reddening}.
 \label{contour_plot_lmc} 
\vspace{\floatsep} 
 \includegraphics[width=0.50\textwidth]{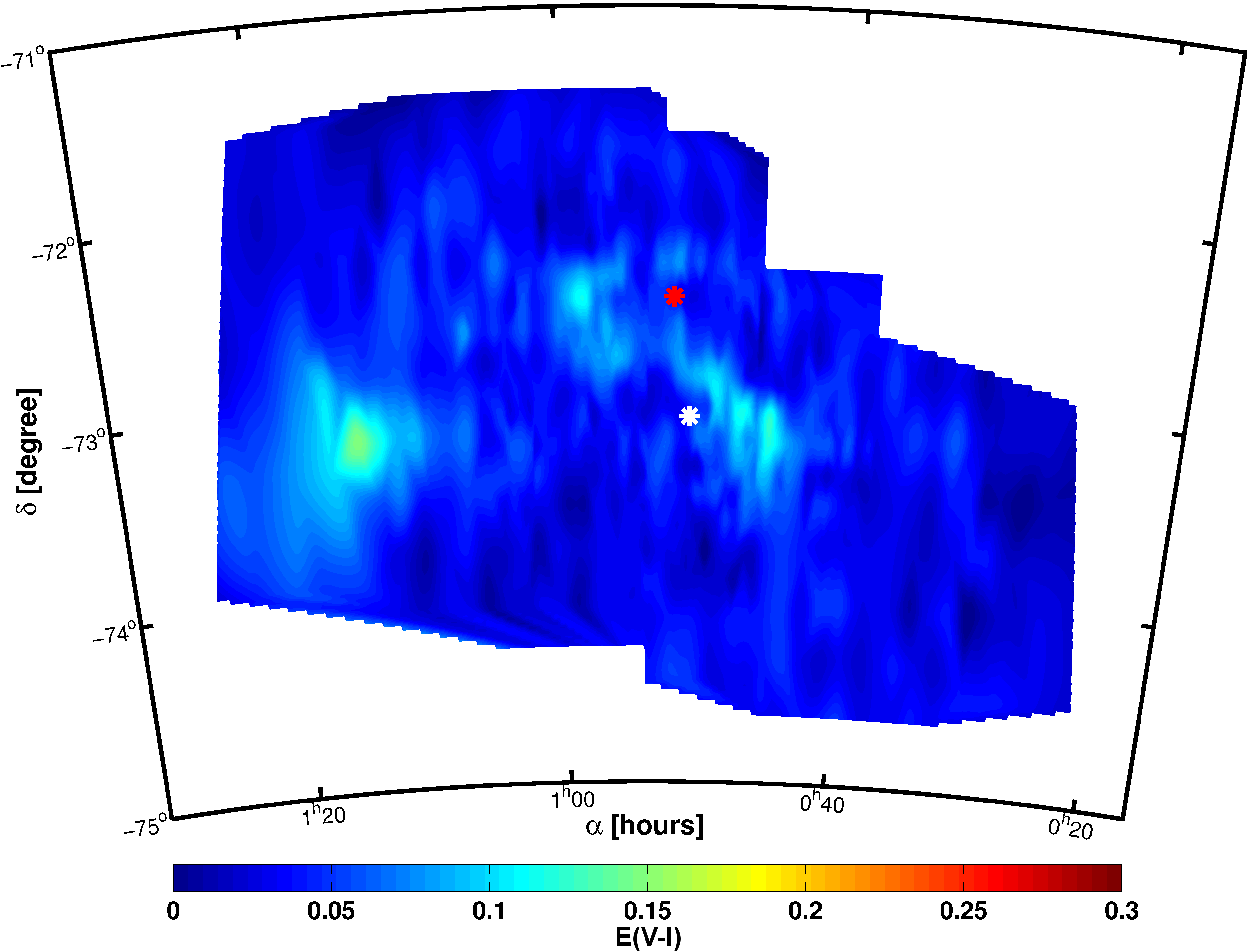}
 \caption{SMC reddening distribution. As before in Figure\,\ref{contour_plot_lmc} low reddening is coded in blue and high reddening in red. The overall reddening is quite low with values around $E(V-I)_{\mathrm{mean}} = 0.04$\,mag. The highest reddening is found along the bar and the wing ($\alpha = 1^{h}15^{m}$ and $\delta = -73^{\circ}10'$) of the SMC. The red star represents the kinematical centre found by \citet{Piatek08}, while the white star shows the centre found from K- and M-dwarfs by \citet{Gonidakis09}.} 
 \label{contour_plot_smc} 
\end{figure}
\begin{figure}
 \centering 
  \includegraphics[width=0.50\textwidth]{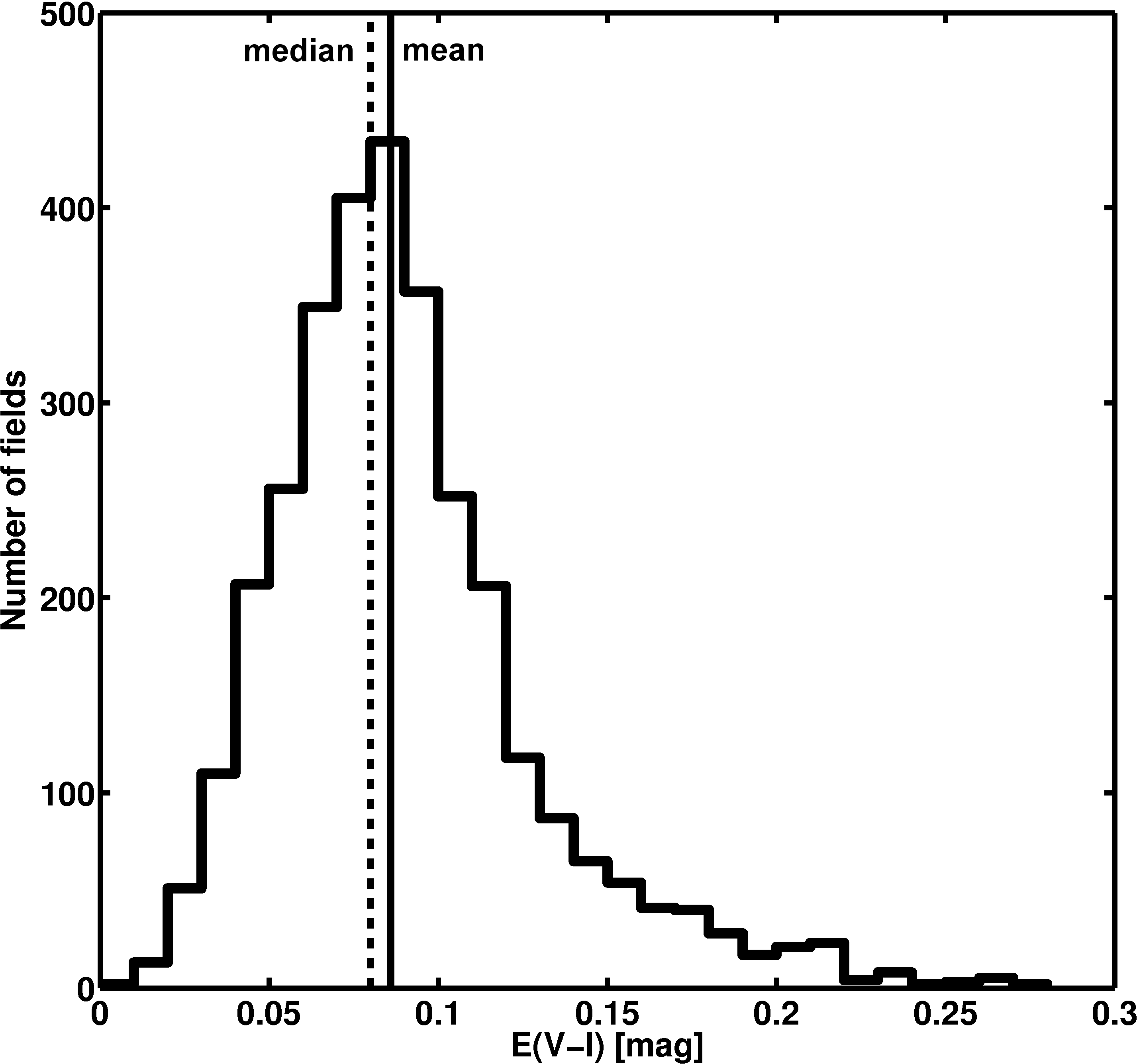}
%
%
  \includegraphics[width=0.50\textwidth]{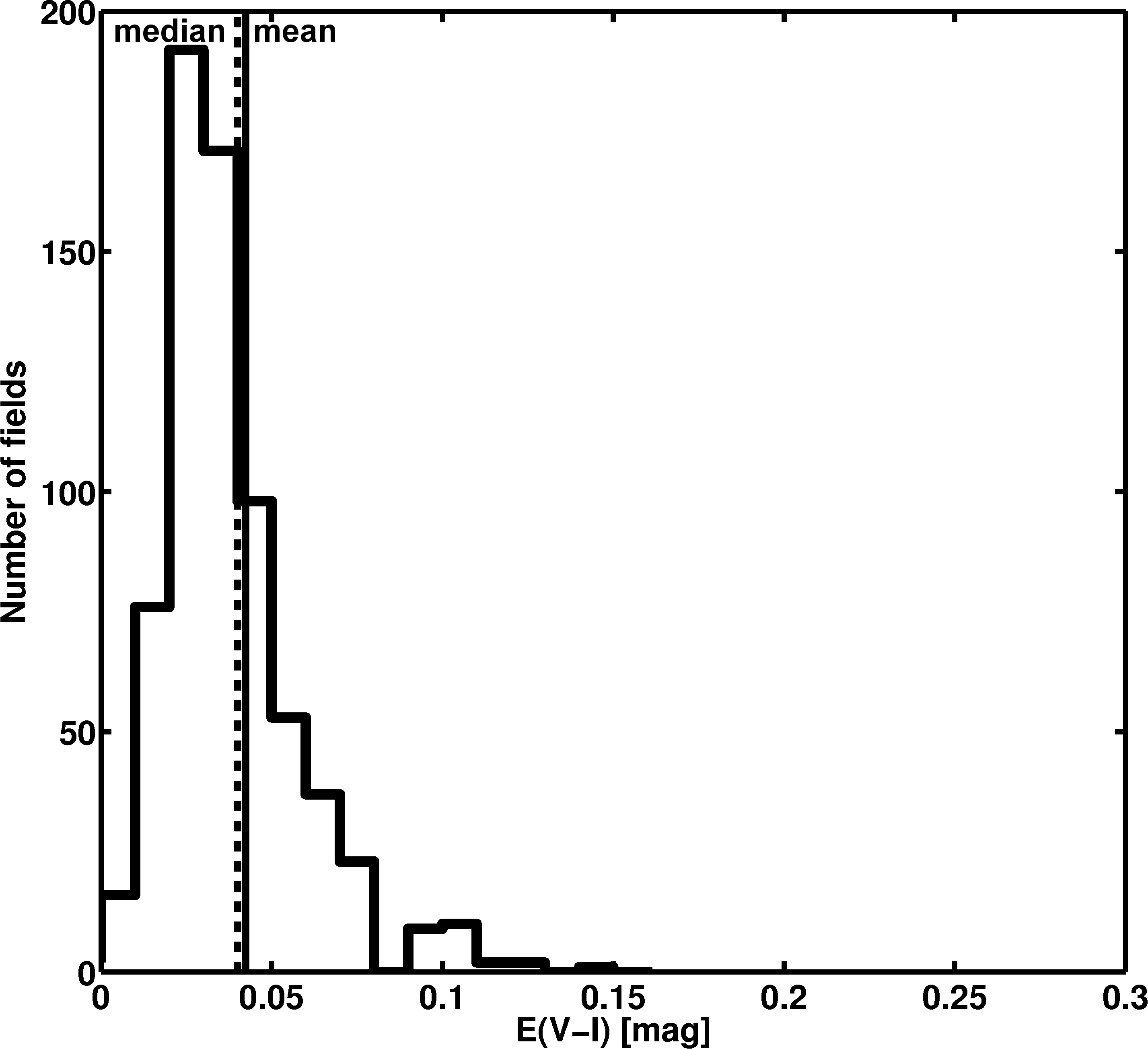}
 \caption{The distribution of reddening $E(V-I)$ based on RC stars for the LMC (top) and SMC (bottom) is shown.  The median and the mean reddening are very similar. We find that the mean reddening values match the peak of the reddening distributions very well.} 
 \label{histogram_distribution_EVI} 
\end{figure}

The unreddened value $(V-I)_0$ was subtracted from the estimated mean RC colours of each of the 3174 subfields in the LMC. The difference corresponds to the mean reddening value $E(V-I)$. These reddening values are plotted in spatially resolved contour maps (Figures\,\ref{contour_plot_lmc} and \ref{contour_plot_smc}). \\\noindent\hspace*{1em}
In the LMC we find a mean reddening of $E(V-I) = 0.09$\,mag and a median value of $E(V-I) = 0.08$\,mag (see Figure\,\ref{histogram_distribution_EVI}). In the central regions, which correspond to the bar, the reddening is quite low. Towards the leading edge of the LMC the reddening increases and reaches its highest values with $E(V-I) = 0.43$\,mag close to the centre of 30\,Doradus. Its centre is located at $\alpha = 5^{h}38^{m}$ and $\delta = -69^{\circ}06'$ \citep{Hog00}. In the southwestern part of the observed field ($\alpha \sim 4^{h}45^{m}$ and $\delta \sim -70^{\circ}20'$) there is a second area with higher reddening visible.  There is only one field with a negative reddening in the dataset. But within the uncertainties this value is consistent with zero. \\\noindent\hspace*{1em}
%
The uncertainty of the reddening values is determined by calculating the $\sigma$ of the gaussian distribution of the stars within the selection box. This represents the broadening of the RC due to metallicity effects, binaries and differential reddening. For the LMC the mean  $1\sigma$ error is $0.07$\,mag. The photometric errors of the observed magnitude of the stars are low due to the large number of repeat observations in the OGLE\,III survey. This leads to a very precise photometry with an average uncertainty of $\sigma = 0.08$\,mag for each star. Each field contains a large number of stars, therefore this error can be neglected. \\\noindent\hspace*{1em} 
Since the central regions contain many more stars than the outskirts of the LMC, the subfields for the determination of the reddening are much smaller than in the outer regions. Therefore, a lot more substructure can be seen here in the reddening map in Figure\,\ref{contour_plot_lmc}.  \\\noindent\hspace*{1em}

\subsubsection*{Regions with high reddening}

\begin{figure}
\centering 
 \includegraphics[width=0.50\textwidth]{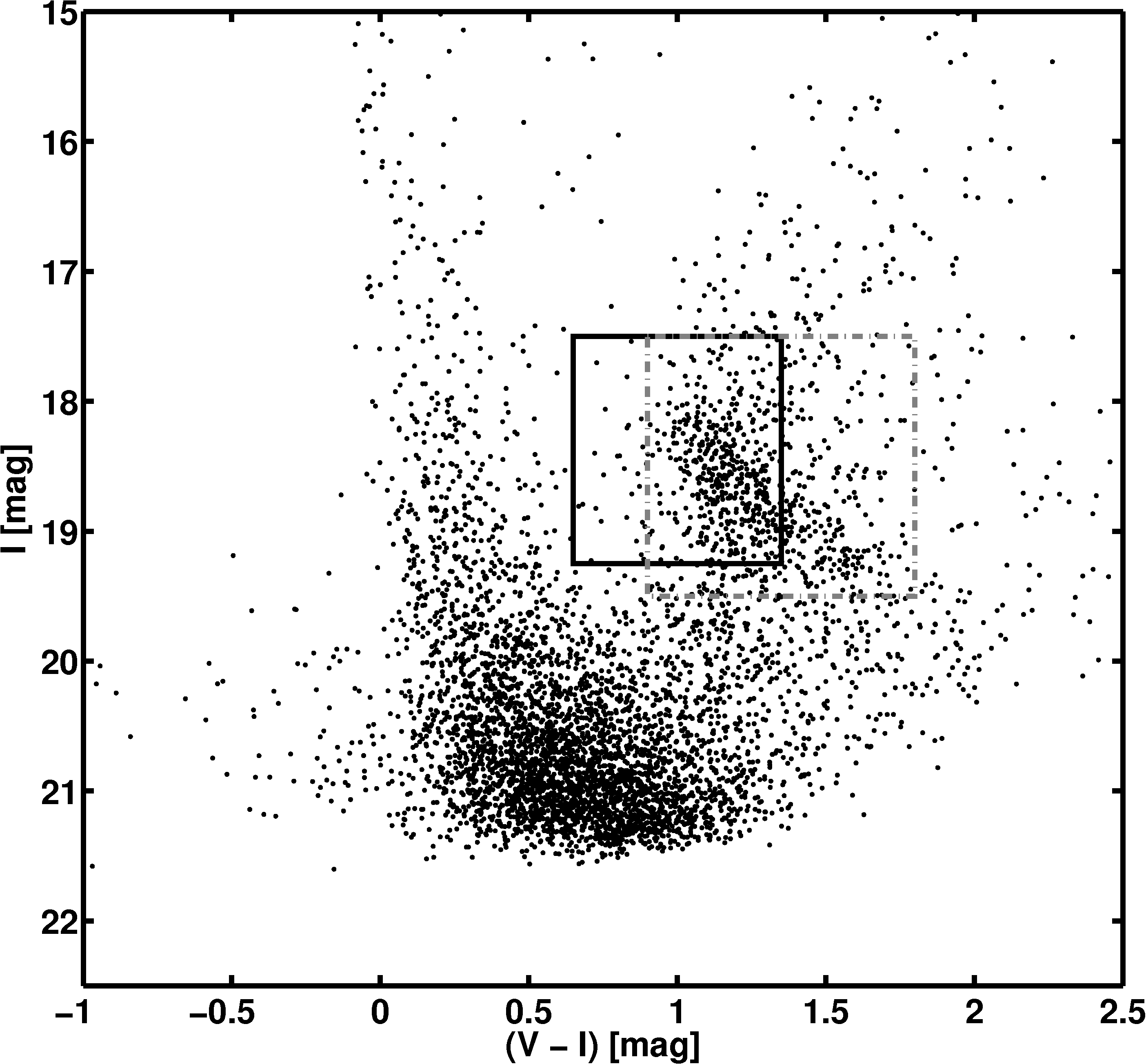}
 \caption{CMD of the LMC subfield with the highest reddening. This field is located in 30\,Doradus at $\alpha = 5^{h}35^{m}$ and $\delta = -69^\circ10'$ with a projected size of $9' \times 9'$. The CMD contains 4386 stars, while 275 stars are in our original selection box, represented by the continuous line. The dashed grey line shows the boundaries of the new box for the RC.} 
 \label{red_clump_in_CMD_high_extinction} 
\end{figure}
\begin{figure}
\centering 
 \includegraphics[width=0.50\textwidth]{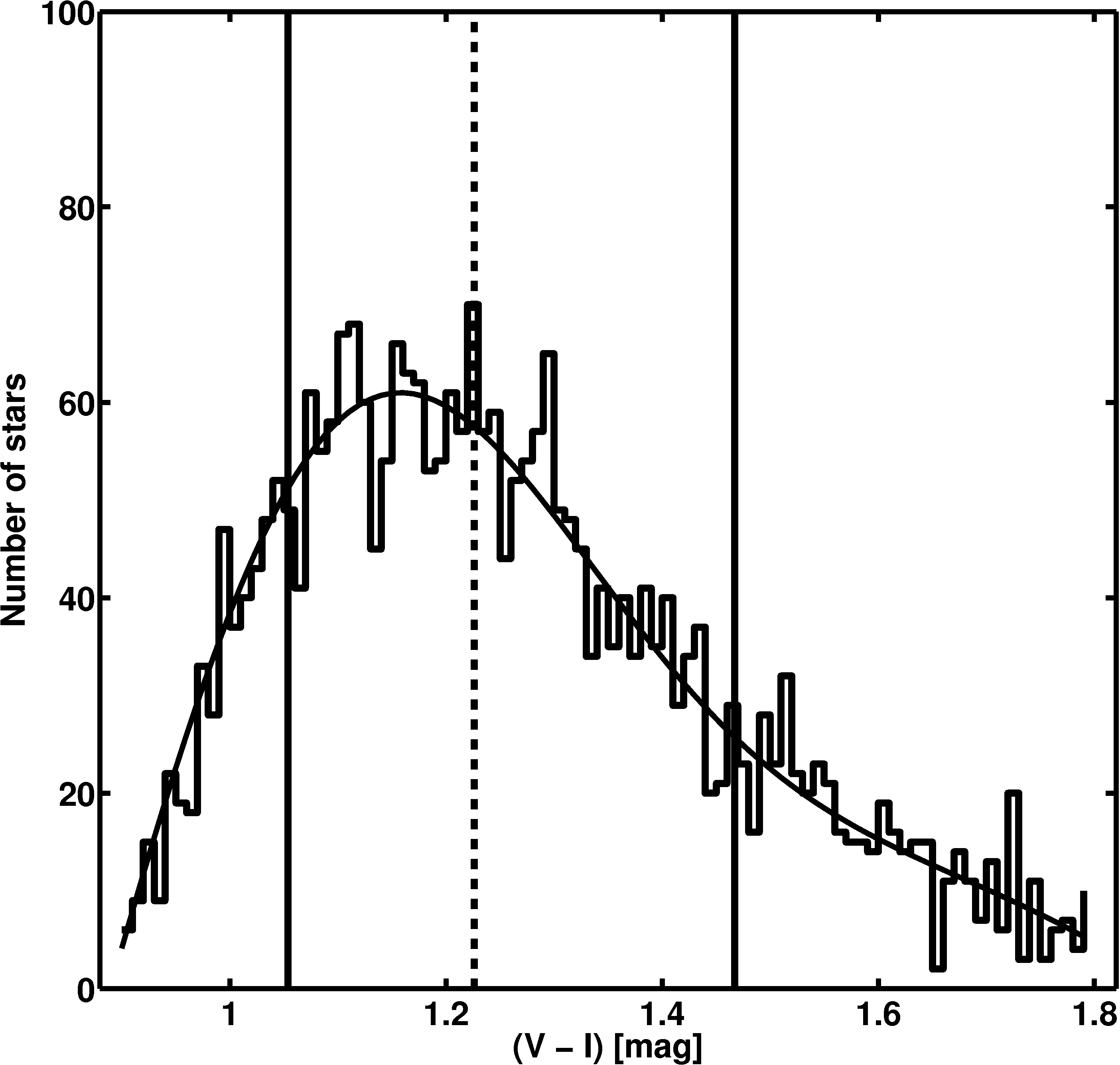}
 \caption{Distribution of reddening for a highly reddened field in the LMC. The two vertical lines show the width that includes 68.2\% of all stars closest to the mean reddening (dashed line) and is considered as the $1\sigma$ width of the distribution here.} 
 \label{histogram_mit_fit_high_extinction} 
\end{figure}

A minor fraction of $2\%$ of the LMC fields show considerable reddening with values of $E(V-I) > 0.2$\,mag. These 60 fields are inspected by eye and only 23 fields with reddening values of $E(V-I) \geq 0.25$\,mag suffer from a very extended red clump, which is caused by differential reddening (Figure\,\ref{red_clump_in_CMD_high_extinction}). The reddening results in these regions are reanalysed by shifting the RC box to $0.9 \leq (V-I) \leq 1.8$\,mag and $17.5 \leq I \leq 19.5$\,mag and refitting the distribution of stars as described above. The results are shown in Table\,\ref{high_reddening_table}. While the peak of the reddening distribution of the field is only marginally shifted, we find a considerable amount of differential reddening for most of the fields. This differential reddening does not follow a Gaussian distribution, but instead typically shows a rapid rise from low reddening to the peak of the distribution, whereas there is a much more shallow and much wider tail towards higher reddening. To determine an estimate for the width of the distribution we calculate the mean of each distribution. To each side of the mean of the RC we include 34.1\% of all stars and determine the colour of the star just outside this border (vertical lines in Figure\,\ref{histogram_mit_fit_high_extinction}). This corresponds to a width of $1\sigma$. \\\noindent\hspace*{1em}
\begin{table}
\begin{center}
\caption{The fields with reddening values of $E(V-I) \geq 0.25$. These fields are reanalysed using a wider RC box and the 1-sigma intervals are calculated.\label{high_reddening_table}}
\begin{tabular}{ccccc}
\tableline\tableline
$\alpha$ & $\delta$ & fieldsize & $E(V-I)_{\mathrm{old}}$ & $E(V-I)_{\mathrm{new}}$ \\    
\tableline                       
82.717 & -68.428 & $9\times9$ & 0.26 & $0.26^{-0.10}_{+0.14}$ \\ 
82.856 & -68.428 & $9\times9$ & 0.35 & $0.35^{-0.17}_{+0.19}$ \\ 
82.995 & -68.428 & $9\times9$ & 0.43 & $0.52^{-0.40}_{+0.14}$ \\ 
83.659 & -69.174 & $9\times9$ & 0.26 & $0.25^{-0.11}_{+0.18}$ \\
83.802 & -69.174 & $9\times9$ & 0.31 & $0.41^{-0.22}_{+0.24}$ \\
83.802 & -69.020 & $9\times9$ & 0.25 & $0.26^{-0.10}_{+0.31}$ \\
83.839 & -69.172 & $9\times9$ & 0.38 & $0.53^{-0.33}_{+0.15}$ \\
83.839 & -69.018 & $9\times9$ & 0.34 & $0.34^{-0.15}_{+0.27}$ \\
83.982 & -69.172 & $9\times9$ & 0.38 & $0.43^{-0.21}_{+0.20}$ \\
83.982 & -69.018 & $9\times9$ & 0.33 & $0.37^{-0.17}_{+0.25}$ \\
84.125 & -69.172 & $9\times9$ & 0.28 & $0.30^{-0.14}_{+0.16}$ \\
84.125 & -69.018 & $9\times9$ & 0.34 & $0.38^{-0.17}_{+0.18}$ \\
84.268 & -69.172 & $9\times9$ & 0.32 & $0.39^{-0.20}_{+0.25}$ \\
84.268 & -69.018 & $9\times9$ & 0.26 & $0.27^{-0.12}_{+0.23}$ \\
85.055 & -69.095 & $18\times18$ & 0.27 & $0.27^{-0.13}_{+0.30}$ \\
84.687 & -69.918 & $9\times9$ & 0.27 & $0.26^{-0.13}_{+0.25}$ \\
84.687 & -69.611 & $9\times9$ & 0.32 & $0.32^{-0.16}_{+0.21}$ \\
84.834 & -69.918 & $9\times9$ & 0.28 & $0.28^{-0.13}_{+0.30}$ \\
84.981 & -69.611 & $9\times9$ & 0.26 & $0.27^{-0.13}_{+0.31}$ \\
84.824 & -70.201 & $9\times9$ & 0.26 & $0.31^{-0.17}_{+0.31}$ \\
84.975 & -70.661 & $9\times9$ & 0.28 & $0.27^{-0.13}_{+0.19}$ \\
85.605 & -71.099 & $9\times9$ & 0.25 & $0.25^{-0.10}_{+0.14}$ \\
86.397 & -69.401 & $18\times18$ & 0.25 & $0.24^{-0.11}_{+0.31}$ \\
\tableline                                  
\end{tabular}
\end{center}
\end{table}

\subsubsection*{Differential Reddening}

For every subfield the mean reddening of all RC stars is measured. Therefore we do not take into account that the dust content may not be evenly distributed within a given subfield. By including $34.1\%$ of all stars on both sides from the maximum, we define the $1\sigma$ width of the distribution, which we use as a measure of the amount of differential reddening. \\\noindent\hspace*{1em}
\begin{figure}
\centering 
 \includegraphics[width=0.50\textwidth]{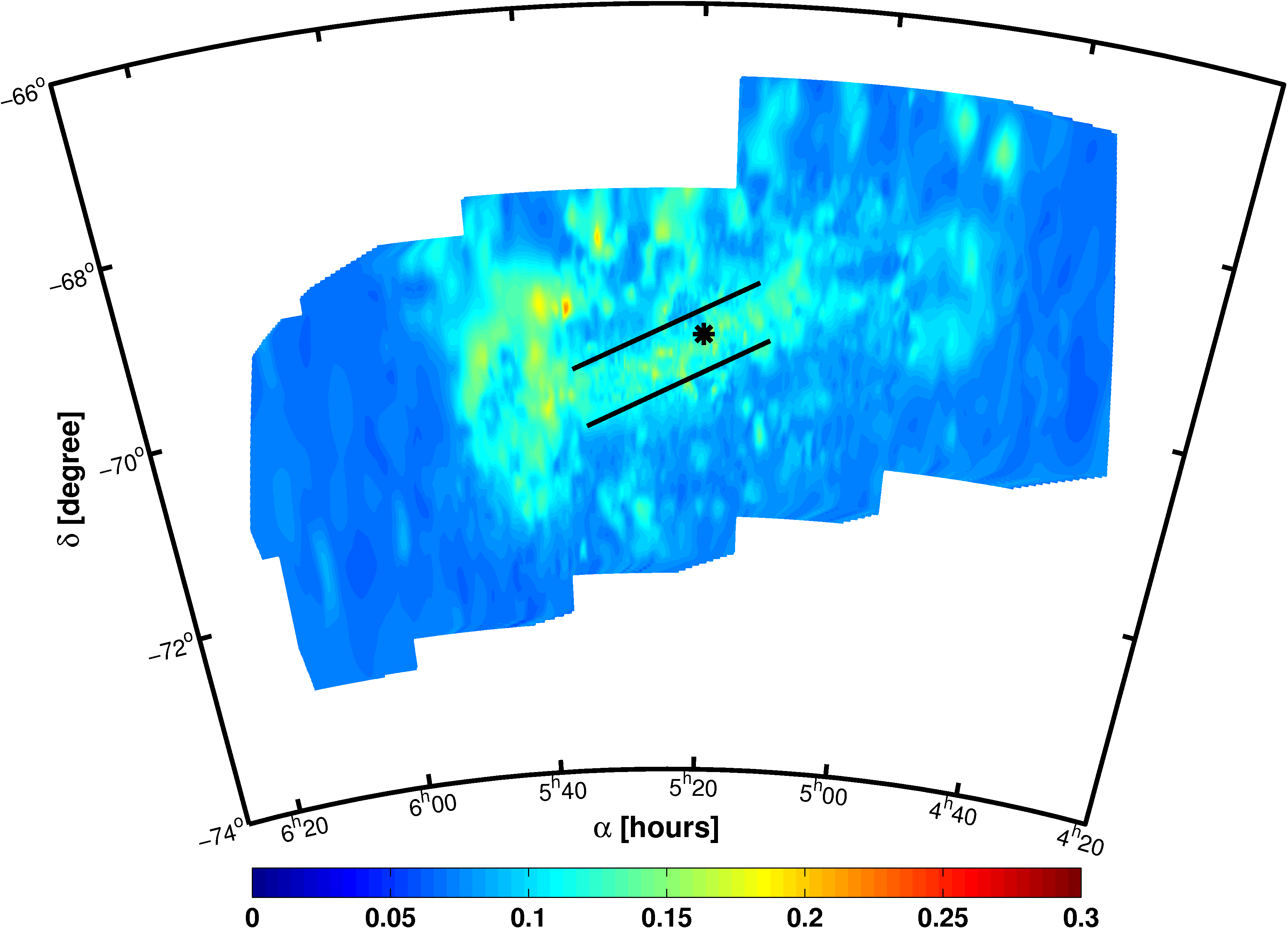}
 \caption{Distribution of differential reddening for the LMC. The differential reddening is high in regions with high reddening, as 30\,Doradus, and in densely populated regions, as the bar.} 
 \label{differential_reddening_RC_LMC} 
\end{figure}
Figure\,\ref{differential_reddening_RC_LMC} shows the distribution of differential reddening in the LMC. The regions with high reddening from Figure\,\ref{contour_plot_lmc} are prominent in the differential reddening map as well. Another region with a considerable spread of the distribution is the bar. Interestingly, the region of increased reddening in the southwestern part of the LMC does not show up as having higher differential reddening.

\subsection{Results for the SMC}
On average the SMC is less reddened than the LMC. For the 693 fields evaluated in the SMC we find the mean and median value to be $E(V-I) = 0.04$\,mag (see Figure \ref{contour_plot_smc} and \ref{histogram_distribution_EVI}). Three main concentrations of higher reddening are revealed in Figure\,\ref{contour_plot_smc}. Two of the concentrations are found at $\alpha = 0^{h}45^{m}$ and $\delta = -73^{\circ}10'$ and at $\alpha = 0^{h}58^{m}$ and $\delta = -72^{\circ}30'$, which coincides with the southwest part and the northern part of the bar, respectively. The overall density of stars is enhanced in these regions. The third area with high reddening is the well known wing of the SMC ($\alpha = 1^{h}15^{m}$ and $\delta = -73^{\circ}10'$). In this region the reddening reaches its maximum with a value of 0.16\,mag. In all regions with high reddening multiple H\,II regions are located \citep{Livanou07}. The lowest calculated reddening is 0.00\,mag. The uncertainty is calculated in the same manner as for the LMC. The same error sources are present and the photometry is as good as for the LMC. We find a mean $1\sigma$ uncertainty of 0.06\,mag.\\\noindent\hspace*{1em}
We measure the differential reddening of the SMC in the same manner as for the LMC. For most fields the differential reddening is quite low and only the bar and the wing show considerable differential reddening. 

\subsection{Data access}
We are making the coordinates of the subfields and the derived RC reddening values for both LMC and SMC available via the German Astrophysical Virtual Observatory (\url{http://dc.zah.uni-heidelberg.de/mcx}). Using the transformation relations published by \citet{Schlegel98} the webpage has the capability to calculate extinctions for 19 different filters in four photometric systems.

\subsection{Use of the RC reddening data}
The reddening maps from the red clump stars cover the complete area of the OGLE\,III survey. They average the reddening within distinct subfields. For all stars in such a field this value can be used as a reliable indication for the average reddening that these stars are experiencing. \\\noindent\hspace*{1em}
Since RC stars are quite red and cool they may not be the first choice for the dereddening of very young and hot stars (see \citet{Grebel95} who discuss the dependence of heterochromatic extinction on stellar parameters such as temperature and \citet{Zaritsky02}, who discuss variations in reddening maps in dependence of stellar population age). Furthermore the reddening in regions with high dust and gas densities, such as star forming regions, might differ significantly with depth. Our method is not able to account for differential reddening along the line of sight, but averages along the sight line, similar as the \citet{Schlegel98} maps do for our Milky Way. \\\noindent\hspace*{1em}
Our RC reddening maps are thus useful for applications in which the bulk properties of stellar populations in the MCs are studied. They are also useful for applications where objects are studied whose properties make the derivation of individual reddening difficult (e.g. individual RC stars or individual red giant branch stars). In studies of individual stellar objects (instead of entire stellar populations), users of the reddening maps should be aware of the afore mentioned caveats of using average (as opposed to individual) reddening values.

%

\section{Reddening based on RR\,Lyrae\,ab stars}
\label{RR_Lyrae}
\subsection{Method}
RR\,Lyrae stars provide an independent means of deriving reddening values. For the MCs the OGLE collaboration presents lightcurves and photometry of RR\,Lyrae covering most of the main bodies of these galaxies \citep{Soszynski09, Soszynski10b}. \citetalias{Pejcha09} used these data to infer a reddening map of the LMC. They assumed one single metallicity for all RR\,Lyrae stars in the LMC, $-1.39$\,dex \citep{Cohen05}, which results in a theoretical intrinsic colour of $(V-I) = 0.48$, with a mean period of 0.57\,days for all RR\,Lyrae in the LMC. However, the spectroscopic studies of RR\,Lyrae stars in the LMC by \citet[e.g.][]{Borissova09} show that a range of metallicities is present. Moreover, \citet{Guldenschuh05} showed in their work that the metallicity of the RR\,Lyrae is negligible only if the colour is measured during minimum light. The OGLE collaboration on the other hand has released mean magnitudes. Hence we decided to extend the work of \citetalias{Pejcha09} by calculating the theoretical value of the colour taking the metallicity of each single star into account. \\\noindent\hspace*{1em}
Using the photometric method established by \citet{Kovacs95, Jurcsik96, Kovacs96} we use the Fourier coefficient $\phi_{31}$ and the period to calculate the metallicity $[Fe/H]$ of each RR\,Lyrae star of type ab. This metallicity is independent of reddening. As shown in the studies quoted above, the V-band light curve of the RR\,Lyrae stars can be decomposed with a Fourier series. The phase parameters can be used to calculate the property $\phi_{31} = \phi_{3} - 3\phi_{1}$. It was shown in the studies cited above that the metallicity can be determined with the knowledge of $\phi_{31}$ and the period. Since the light curve properties are different in different photometric bands a separate calibration of $\phi_{31}$ and the period has to be carried out if the data are not taken in the V-band. This is relevant for the OGLE data, where the light curves are best sampled in the I-band. \citet{Smolec05} provides a relation valid when $\phi_{31}$ is computed from the I-band instead of V-band magnitudes. We use Smolec's relation to infer the metallicities of the RR\,Lyrae from OGLE's I-band $\phi_{31}$. Then we use the metallicity dependent relations for the V- and I-band, found by \citet{Catelan04} by theoretical modelling of lightcurves, to infer the absolute colour $(V-I)_0$ of each star. Further details on the results for the RR\,Lyrae\,ab metallicities are presented in a separate paper \citep{Haschke11_MDF}.\\\noindent\hspace*{1em}
The measurements of the RR\,Lyrae have the same photometric uncertainties as the RC stars: $\sigma = 0.08$\,mag. For the observed colour this is the only source of uncertainty taken into account. In the I-band the absolute magnitude depends on the period and on the metallicity of the RR\,Lyrae star, while the V-band magnitude only depends on metallicity. For the metallicity derived via Fourier decomposition \citet{Smolec05} states an uncertainty of $\sigma = 0.18$\,dex, while the uncertainty of the periods, measured by the OGLE collaboration, is negligible. The uncertainty of the reddening is therefore determined to be $\sigma = 0.06$\,mag. Many relations for calculating the absolute magnitudes of RR\,Lyrae stars have been published \citep[see][for a review]{Sandage06}.\\\noindent\hspace*{1em}
\citetalias{Pejcha09} found that imposing certain colour and magnitude constraints ensures a reliable selection of RR\,Lyrae stars of type ab that are likely members of the LMC. We adopt similar constraints: periods between 0.45\,days $<$ P $<$ 0.80\,days, amplitudes between 0.30\,mag $< \Delta\,I <$ 0.85\,mag, observed magnitudes of V $>$ 18\,mag, observed colours larger than $(V-I) > 0.40$\,mag and metallicities between $-2.0 < [Fe/H] < 0.0$. The 17693 RR\,Lyrae\,ab found in the LMC are reduced to 13490 stars. In the SMC the number of stars is reduced from 1864 to 1529 RR\,Lyrae\,ab. \\\noindent\hspace*{1em}

\subsection{Results for the LMC}
\begin{figure}
 \centering 
 \includegraphics[width=0.50\textwidth]{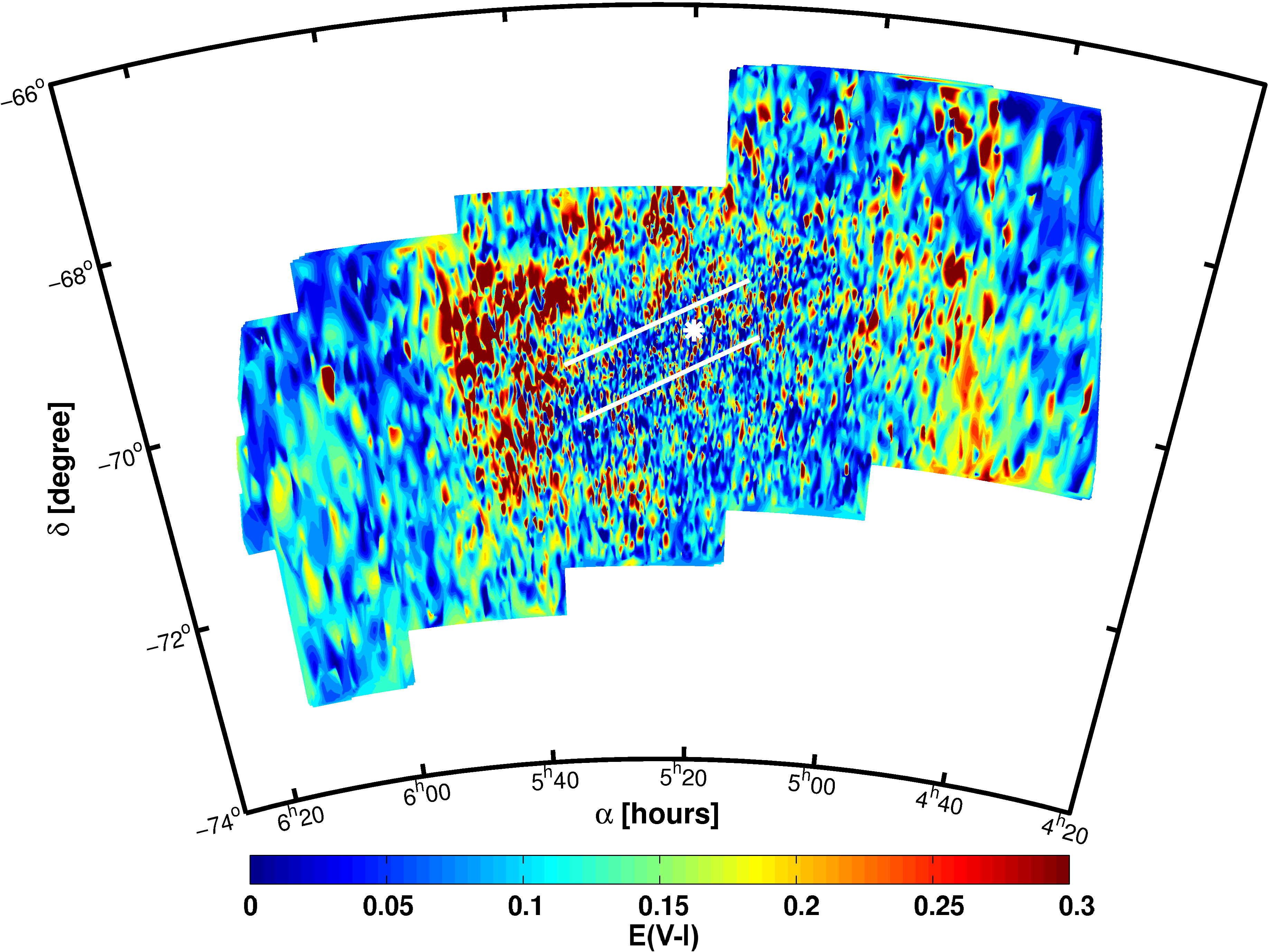}
 \caption{A reddening map for the LMC based on individual RR\,Lyrae stars, taking their metallicity into account. The map reveals large differences between stars that are in projection very close to each other on the sky. These small-scale variations appear to be due to differences in the line-of-sight extinction. } 
 \label{reddening_RR_LMC} 
\vspace{\floatsep} 
 \includegraphics[width=0.50\textwidth]{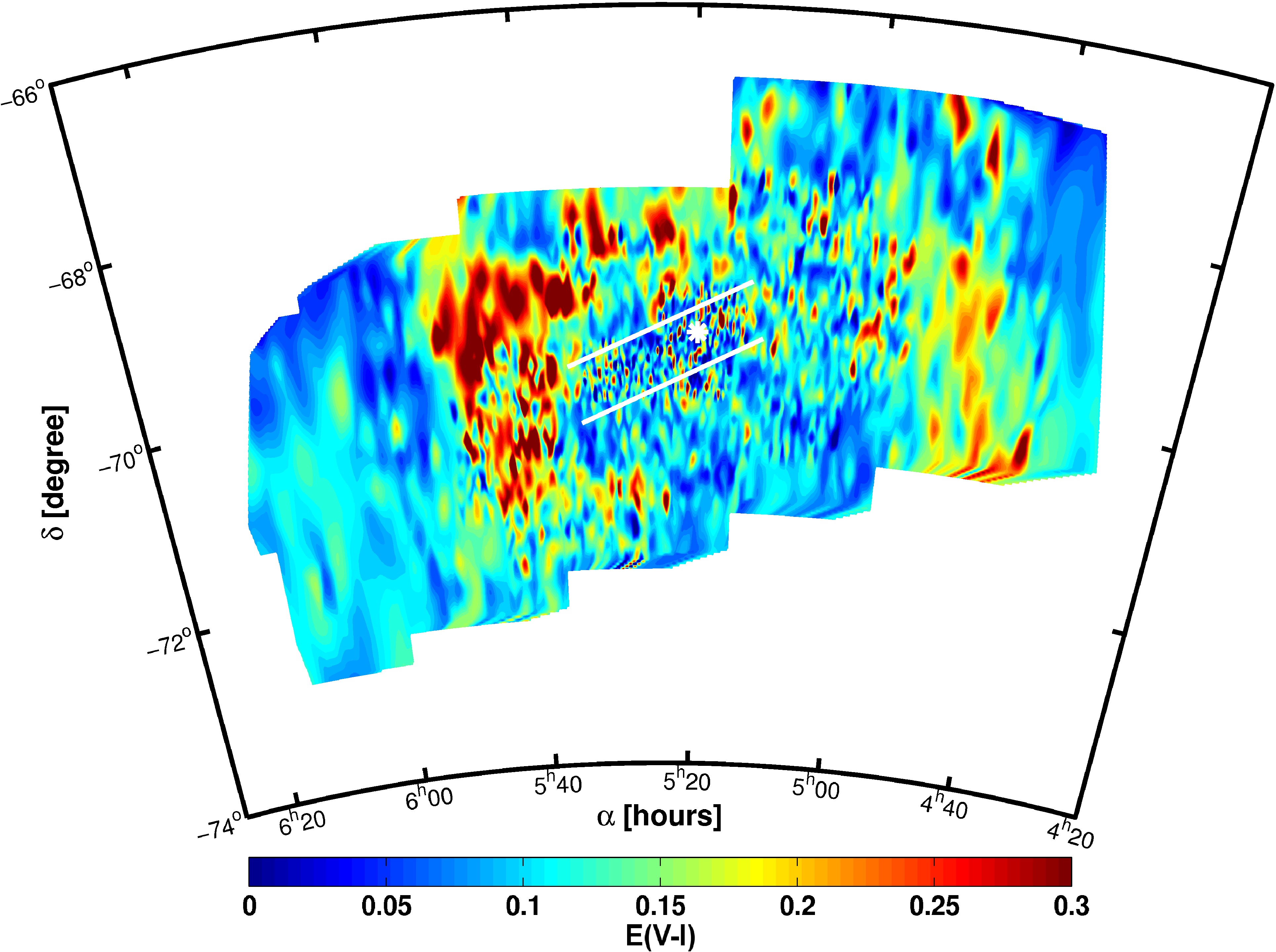}
 \caption{A reddening map of the LMC based on RR\,Lyrae averaged across the same fields as used for the RC method. A slightly higher mean reddening is visible than for the RC reddening, but overall the map shows a similar distribution of the reddening as in Figure \ref{contour_plot_lmc}, even though the methods are completely independent.} 
 \label{reddening_RR_mean_LMC} 
\end{figure}

For each of the 13490 remaining LMC RR\,Lyrae a reddening value is calculated. The resulting values are plotted in Figure \ref{reddening_RR_LMC}. To reduce the fluctuations due to the varying reddening values of the individual RR\,Lyrae stars we obtain a new reddening map of the LMC by averaging over the individual RR\,Lyrae reddening values in the same subfields chosen for the RC reddening. The mean reddening of all RR\,Lyrae stars located in each field is adopted in the reddening map shown in Figure \ref{reddening_RR_mean_LMC}. 120 fields are returned with no star allocation since the RR\,Lyrae are much more sparsely distributed than the RC stars. For each of these empty fields the mean reddening of the neighbouring fields is assigned as reddening. \\\noindent\hspace*{1em}
The reddening map of the LMC obtained from RR\,Lyrae stars reveals a lot of structure. The highest reddening values are obtained in the leading edge, containing the spur and 30\,Doradus. The highest value of a single RR\,Lyrae star is $E(V-I) = 1.42$\,mag, while seven stars have a reddening of $E(V-I) > 1.0$\,mag. The highest averaged field reddening value is $E(V-I) = 0.66$\,mag. Another region with high RR\,Lyrae reddening values is found in the southwestern part of the OGLE\,III field ($\alpha = 4^{h}45^{m}$ and $\delta = -70^{\circ}20'$). In the bar region the reddening is quite low and overall we find a mean reddening of 0.13\,mag for the individual stars and the fields. The median reddening is slightly lower, namely $E(V-I) = 0.11$\,mag. We find a standard deviation of 0.06\,mag,  a value of the same order of magnitude as the error for the RC method. While Figure\,\ref{reddening_RR_mean_LMC} contains only a few spots of high reddening fluctuations between neighbouring areas, these are much more striking in Figure\,\ref{reddening_RR_LMC}, which is to be expected since we are considering individual stars here. These fluctuations may be indicative of differential reddening along the line of sight. \\\noindent\hspace*{1em}
Overall we find 635 stars (which corresponds to 5\% of the whole sample) with a negative reddening down to $E(V-I) = -0.20$\,mag. These lead to 33 fields (1\% of all fields) with negative values. The field with the lowest mean RR\,Lyrae reddening $E(V-I)$ has $-0.11$\,mag. The mean for the individual stars as well as for the fields with these negative values is $E(V-I) = -0.03$\,mag. We inspect the lightcurve of every star with negative reddening. As one would expect, the shape of the light curves differs from star to star, but is well within the expected variations. The mean period of the RR\,Lyrae with negative reddening is higher by $10\%$ than that of the bulk of the RR\,Lyrae stars. With a mean of $[Fe/H]_{E(V-I)<0} = -1.40$ the metallicity of these stars is lower than the overall mean of $[Fe/H]_{\mathrm{mean}} = -1.23$. But there is no clear trend in the sense that most low metallicity stars or most high period stars have too low a reddening. The details of the metallicity distribution function will be discussed in a separate paper. The origin of the negative reddening remains unclear. If we set all negative values to zero, the resulting mean reddening of the LMC increases by just $1.5\%$.\\\noindent\hspace*{1em}

\subsection{Results for the SMC}

\begin{figure}
 \includegraphics[width=0.50\textwidth]{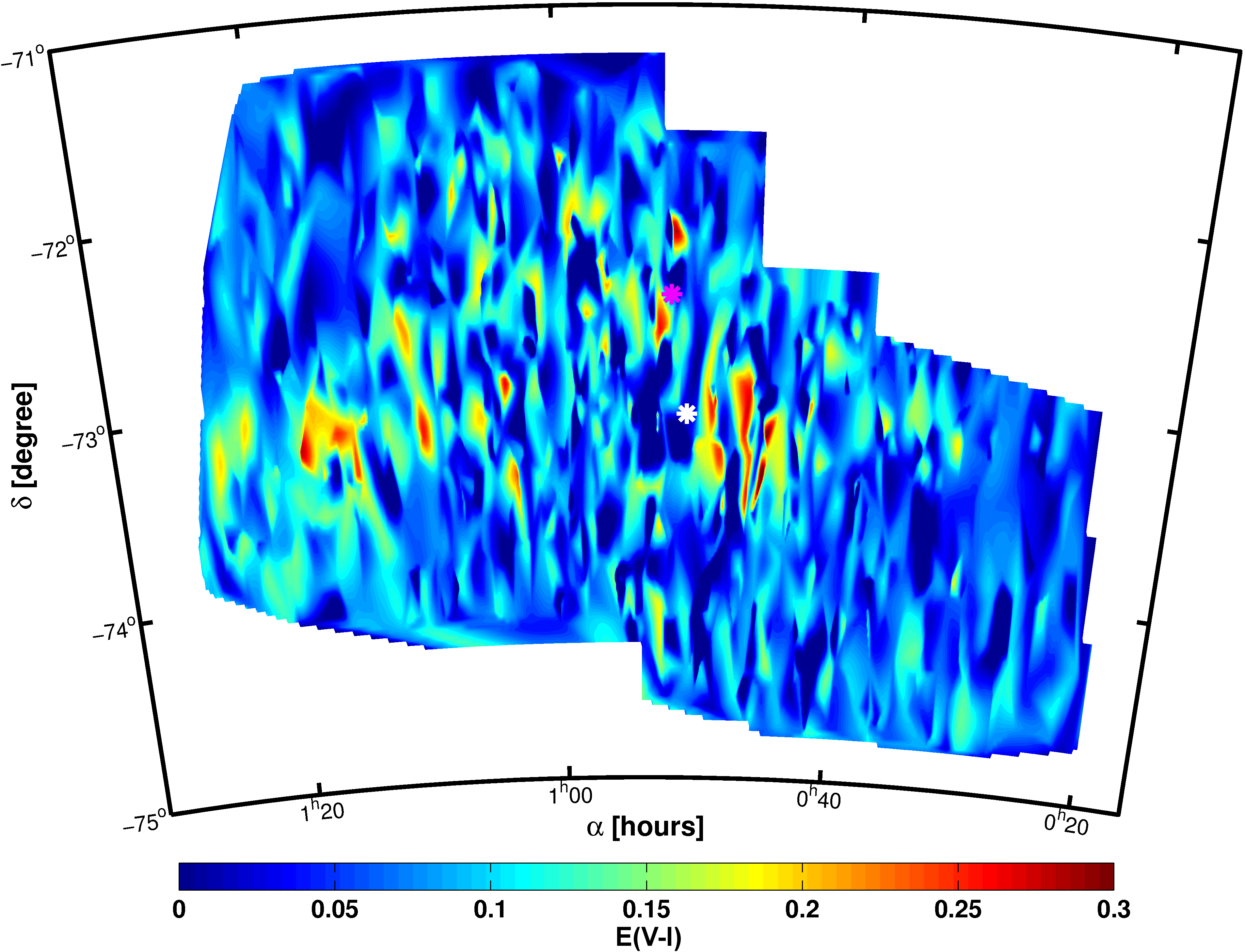}
 \caption{A reddening map for the SMC based on reddening estimates for individual RR\,Lyrae stars. Due to differential reddening effects the map shows a much patchier behaviour than Figure\,\ref{contour_plot_smc}. } 
 \label{reddening_RR_SMC} 
\end{figure}

Using the same procedure, we calculate reddening values for the SMC 1529 RR\,Lyrae\,ab stars. The resulting reddening is shown in Figure\,\ref{reddening_RR_SMC}. \\\noindent\hspace*{1em}
Even though much more substructure is visible in Figure\,\ref{reddening_RR_SMC} than in Figure\,\ref{contour_plot_smc} the overall patterns are similar. The bar and the wing have the highest reddening values. The differences on small scales are, as for the LMC, interpreted as differential reddening. The highest reddening value in the SMC with $E(V-I) = 0.36$\,mag is located in the southwestern bar region. Overall we find that the mean reddening has a value of $E(V-I) = 0.07$\,mag, with a standard deviation of 0.06\,mag. \\\noindent\hspace*{1em}
Negative reddening is found for 114 stars or 7\% of the sample. The lowest reddening for an individual star is  $E(V-I) = -0.15$\,mag, while the mean of all stars with negative reddening is $E(V-I) = -0.03$\,mag. If all negative reddening values are set to 0 the mean reddening increases by $3.5\%$. The mean metallicity of the stars with negative reddening is lower than for the whole sample of stars, i.e. we see the same trend as in the LMC. While the entire sample has a mean metallicity of $[Fe/H]_{\mathrm{mean}} = -1.41$, for the negative reddening stars a metallicity of $[Fe/H]_{E(V-I)<0} = -1.63$ is obtained. The details of the metallicity distribution will be discussed in a separate paper. The period of the RR\,Lyrae stars with negative reddening is increased by 5\% compared to the mean of all stars. Nevertheless, a clear trend that most of the metal-poor stars have a lower reddening is not apparent. \\\noindent\hspace*{1em}
The intrinsic uncertainties of the parameters used for the reddening estimates are the same in the SMC as in the LMC. The uncertainty of the SMC RR\,Lyrae reddening is calculated to be $\sigma = 0.06$\,mag. This is comparable with the uncertainty of the RC method.

\subsection{Use of the RR\,Lyrae reddening data}
Using the RR\,Lyrae stars of the MCs we determine reddening values for lines of sight to individual stars. These values are strictly only valid for these particular RR\,Lyrae stars. We are not able to constrain the stars' distance independently of the reddening measurement. Therefore we are not able to distinguish between objects located at varying depth that might be affected by different amounts of differential reddening. \\\noindent\hspace*{1em}
The RR Lyrae reddening maps are not necessarily useful for the reddening correction of other types of stars close by in projection. Therefore we do not provide the reddening values via the virtual observatory. Nonetheless, these data provide a detailed account of the reddening experienced by the old population of the MCs and are interesting to compare to reddening maps presented in \citet{Zaritsky02, Zaritsky04} as well as to our RC reddening maps. We can obtain insights into the differential reddening which is of interest for a broad variety of subjects such as, e.g., gravitational lensing. Moreover these data are useful for comparison with the structural and geometrical properties of the MCs.

%

\section{Discussion}
\label{discussion}
\subsection{The zeropoint of the RC colour}
\label{RC_colour}

The crucial point in our RC reddening determinations are the unreddened theoretical colours, which depend on the metallicity and the age of the stellar population (see Section\,\ref{Caveats}). In \citetalias{Subramaniam05} and in our work the value for the RC colour was adopted from the work of \citet{Girardi01}. Both studies assume a metallicity of $z=0.004$, which corresponds to a RC colour between 0.90\,mag and 0.95\,mag for stars with an age of $2-12$\,Gyr. Other works, such as \cite{Cole05} with spectroscopic measurements of red giants, find similar metallicities as assumed in these studies. Based on the photometric measurements of RC stars by \citet{Olsen02} we adopt the value $(V-I)_0 = 0.92$. \\\noindent\hspace*{1em} 
The reddening values obtained by \citet{Udalski99a, Udalski99b} with the RC method are much higher ($E(V-I)_{\mathrm{mean}} = 0.20$\,mag) than the reddening in \citetalias{Subramaniam05} and in this study. \citetalias{Subramaniam05} suggest that this discrepancy may be resolved by adopting a much lower metallicity of $z = 0.001$ for the LMC RC population, which would result in a bluer colour for the RC. Since neither the concrete method, nor the adopted metallicity is mentioned in \citet{Udalski99b} we can only speculate about the cause(s) of the differences. \\\noindent\hspace*{1em} 
The overall metallicity of the SMC is lower than that of the LMC. \citet{Matteucci02} showed that the red clump in the SMC is shifted to bluer colours as compared to the LMC. \citet{Girardi01} provide a RC colour of $0.90 < (V-I)_0 < 0.95$\,mag for $z=0.004$, while their models lead to a colour between $(V-I)_0 = 0.78$\,mag and $(V-I)_0 = 0.84$\,mag for $z=0.001$. We assume that the metallicity of the SMC RC lies a bit below the metallicity value of the LMC and choose a RC colour of $(V-I)_0 = 0.89$\,mag. The metallicities of the intermediate age populations show a considerable spread at any given age \citep{Glatt08b}. The mean metallicities found by \citet{Cole98} and \citet{Glatt08b} agree with our assumption of a metallicity $z$ between $0.002$ and $0.003$.

\subsection{Possible caveats of the methods used}
\label{Caveats}
For the RC method different effects lead to a broadening of the clump, such as metallicity and age spreads, binaries, and differential reddening. \\\noindent\hspace*{1em} 
A lower metallicity would lead to a bluer colour of the RC and therefore enhance the overall reddening of all fields. The existing measurements of the metallicity of RC stars make this effect unlikely. An age spread among RC stars is present in all fields, but the effect on the colour is quite small as long as the metallicity does not change dramatically.\\\noindent\hspace*{1em} 
For the RR\,Lyrae stars we have derived the metallicity of each star individually in a separate paper \citep{Haschke11_MDF}. We take these different metallicities into account while calculating the absolute colours of each RR\,Lyrae star. The uncertainties of the metallicity are about 0.2\,dex, which corresponds to a colour difference of $\Delta(V-I)_0 = 0.001$\,mag.\\\noindent\hspace*{1em} 
The RR\,Lyrae stars are up to a few Gyr older than the RC stars, hotter by about 2000K \citep{Puzeras10, Smith04}, and bluer by about $(V-I) = 0.3$\,mag. The absolute magnitude of RR\,Lyrae and RC stars are similar. As discussed in Section\,\ref{RC_colour_two} these differences are not expected to have a significant effect on the reddening obtained for these stars. Assuming that the spatial distribution of RC and RR\,Lyrae stars is similar, differences are mainly expected due to differential reddening, which gets averaged out in the RC method, while fully affecting the individual RR\,Lyrae measurements. 
\subsection{Assumptions on the Geometry}
\label{Geometry}
If we make the simplified assumption that the dust forms a thin, equidistant sheet across the LMC (or SMC), then the amount of reddening suffered by a particular region can be directly translated into the geometry of that region:  The more reddened a region is, the farther it is behind the sheet; the less reddened it is the more stars are in front of the sheet.  However, this reasoning becomes more difficult if the dust is not evenly distributed (as we know it isn't; see, e.g., the Spitzer dust maps).  Moreover, the dust properties vary across the face of the Magellanic Clouds as traced by, e.g., the varying appearance at 8 microns and 160 microns.  Ideally, one would like to use an independent distance determination method in order to infer the 3-D geometry of the Clouds.  We attempt to do this in a forthcoming paper.
\subsection{Comparison of our two approaches}
\label{RC_colour_two}
\subsubsection*{LMC}
\begin{figure}
 \includegraphics[width=0.50\textwidth]{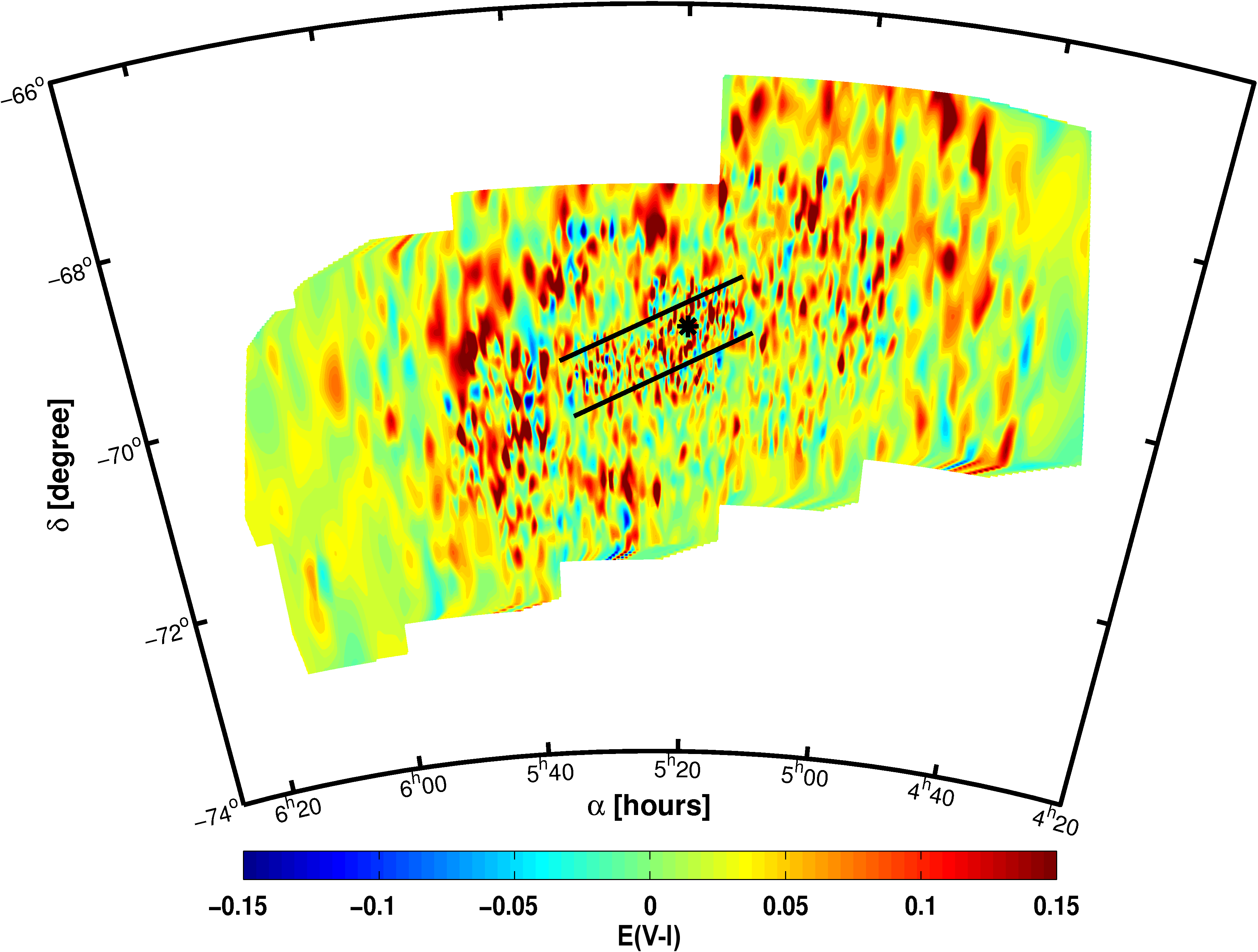}
 \caption{Difference map of the LMC. The red clump reddening is subtracted from the RR\,Lyrae values: $E(V-I)_{RR} - E(V-I)_{RC}$. A small trend of higher reddening estimated from RR\,Lyrae stars is apparent. We find a mean difference of $0.03$\,mag, smaller than the mean error of both methods. Depending on the location the difference ranges from $-0.19$\,mag to $+0.52$\,mag.} 
 \label{comparison_RR_RC_LMC} 
\end{figure}

In order to compare the RR\,Lyrae reddening with the RC reddening, the same field selection as in Section \ref{red_clump} is used in the LMC to find average values of RR\,Lyrae reddening. \\\noindent\hspace*{1em}
As shown by \citet{Grebel95}, reddening varies as a function of temperature (or colour). However, in the colour range of RC stars and RR\,Lyrae hardly any variations are to be expected (see, e.g., Fig.3 in the above reference). Hence, theoretically both the RC and the RR\,Lyrae approach should yield fairly similar results if the reddening suffered by both types of stars is comparable. \\\noindent\hspace*{1em}
For the LMC the overall agreement between the RC and the RR\,Lyrae map is striking, although the northwestern regions ($\alpha \sim 5^{h}00^{m}$ and $\delta \sim -68^{\circ}00'$) have on average a higher reddening in the RR\,Lyrae map (Figure\,\ref{comparison_RR_RC_LMC}). In general we find a mean difference between the mean RC and the mean RR\,Lyrae reddening of $E(V-I) = +0.03$\,mag when subtracting the RC map from the mean reddening map of the RR\,Lyrae. This is a difference of less than 1$\sigma$. We find a few regions with reddening values up to $0.52$\,mag higher in the RR\,Lyrae determination as compared to the RC map at the same location. Some fields have higher reddening values from the RC method than from the RR\,Lyrae estimates. The field with the highest difference is found with $E(V-I)_{RC} - E(V-I)_{RR} = 0.19$\,mag. But in general only very few fields deviate strongly from each other. In only about $8\%$ of the fields the variation exceeds a difference of more than $0.12$\,mag or $2\sigma$. We attribute these differences primarily to differential reddening and small number statistics affecting the RR\,Lyrae values. \\\noindent\hspace*{1em}
At the spatial position of 30\,Doradus the reddening is highly dependent on the location of a given star with respect to the star-forming region. \citet{Westerlund97} and \citet{Marel06} mention that 30\,Doradus seems to be located on the far side of the LMC as seen from us. If that were true the dust and gas of 30\,Doradus would have an effect on only a minority of the much older RC stars. Therefore the (mean) reddening estimate of this region would be biased towards lower values and a long tail of higher reddening values would be measured (see Figure\,\ref{histogram_mit_fit_high_extinction}). For the RR\,Lyrae stars we compute a reddening value for every star. This results in a substantial differential reddening, due to the differing distances of these stars, and in a spread of the estimated reddening values in each subfield. Averaging these values for a given subfield does not reduce the contribution of highly reddened stars behind 30\,Doradus, as much as it might be the case for the RC method. We obtain higher average values for the same subfield by using the RR\,Lyrae stars than for the RC stars, which leads to the assumption of 30\,Doradus being at the far side of the LMC. \\\noindent\hspace*{1em}
%
The RR Lyrae stars are located at different distances throughout the LMC. In addition the dust content in the body of the LMC leads to differential reddening. In Figure\,\ref{reddening_RR_LMC} the effects of differential reddening are obvious. Fields very close to each other may have very different reddening values. For 958 stars, or 7\%, of all RR\,Lyrae stars the reddening values of the individual stars differ by more than $0.18$\,mag or $3\sigma$, from the corresponding mean value of the RC reddening. The effect of differential reddening therefore affects only a minor number of stars.  \\\noindent\hspace*{1em}
The adopted mean colour of the RC is crucial for the reddening computation. The very good overall agreement of RC and RR\,Lyrae maps in areas with very low reddening leads us to the conclusion that the theoretical colours are well chosen. For regions with $E(V-I)_{RR} < 0.1$\,mag a mean difference of only 0.004\,mag is found calculating $E(V-I)_{RR} - E(V-I)_{RC}$. \\\noindent\hspace*{1em}
\subsubsection*{SMC}
\begin{figure}
 \includegraphics[width=0.50\textwidth]{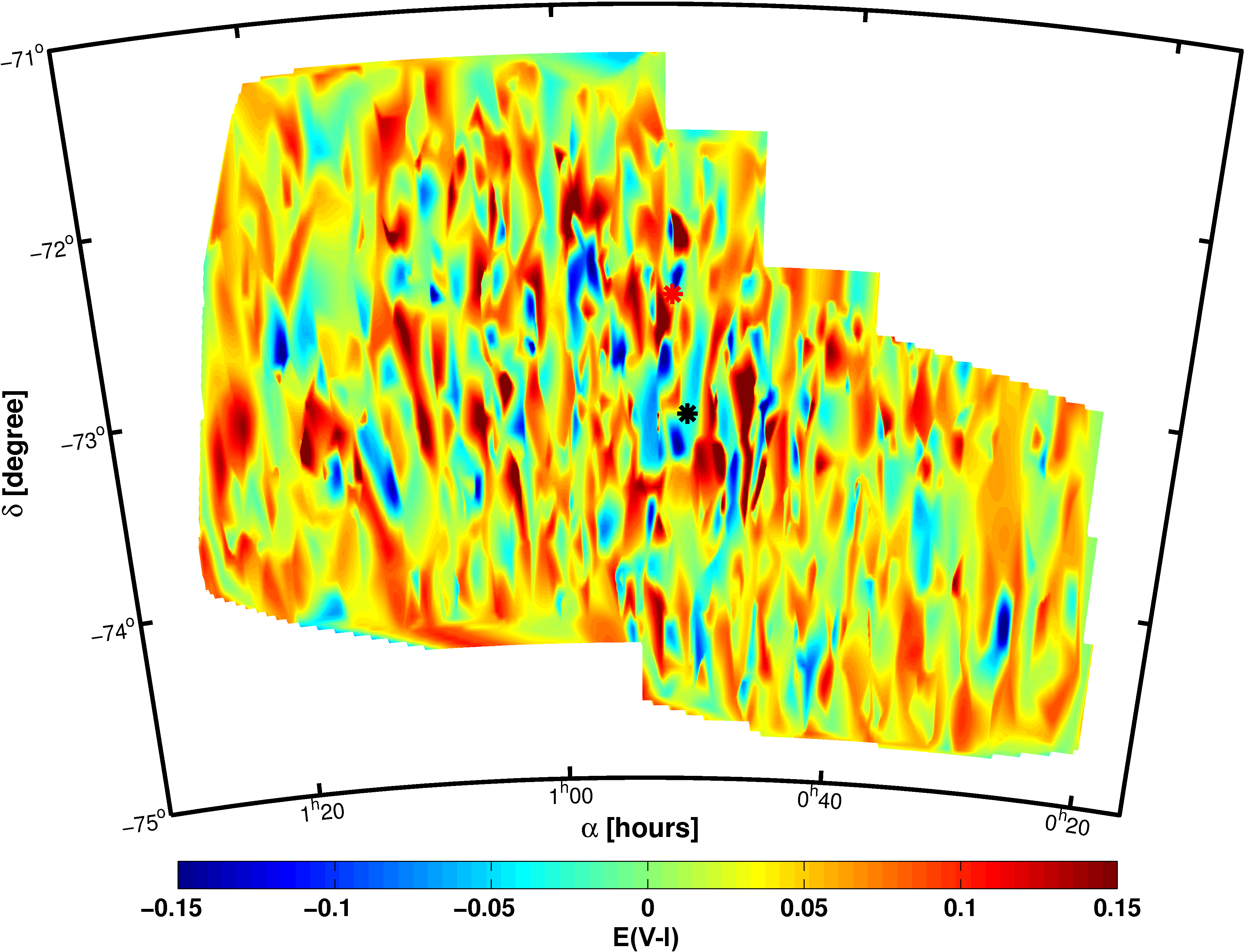}
 \caption{Same as Figure\,\ref{comparison_RR_RC_LMC} for the SMC. The RR\,Lyrae map is much more patchy than the RC map. We conclude that this is a sampling effect owing to the small number of RR\,Lyrae. For regions with very low reddening the discrepancy between the two methods is basically zero.} 
 \label{comparison_RR_RC_SMC} 
\end{figure}

The SMC is populated much more sparsely in RR\,Lyrae stars, thus many fields defined earlier when using the RC method contain less than three RR\,Lyrae stars. Therefore we decided to compare the individual values of the RR\,Lyrae stars with the RC reddening (see Figure\,\ref{comparison_RR_RC_SMC}). \\\noindent\hspace*{1em}
On average the RR\,Lyrae stars have a reddening that is higher by $E(V-I) = +0.03$ than the corresponding RC field. The differences between the two approaches vary between $-0.19$\,mag and $+0.28$\,mag. 348 stars (23\%) have lower reddening with the RR\,Lyrae method than the corresponding RC reddening value. Only 48 stars, or 3\%, of the RR\,Lyrae deviate by more than $2\sigma$ from the mean reddening of the field obtained by the RC method. \\\noindent\hspace*{1em}
Even though the mean differences are small Figure\,\ref{comparison_RR_RC_SMC} reveals that the RR\,Lyrae reddening map is much more patchy than the very smooth RC map. While we inspect and average over fields of a certain size with the RC method, the RR\,Lyrae measure just the line of sight reddening of individual objects. We find only 13 stars, which is less than $1\%$ of the sample, where the reddening value of the RR\,Lyrae deviates by more than $3\sigma$, or $0.18$\,mag, from the average value of the corresponding RC field. Hence we conclude that the poor sampling of the field is the major contribution for this patchy behaviour of the RR\,Lyrae reddening map. \\\noindent\hspace*{1em}
The mean colour of the RC method for the SMC was adopted from \citet{Udalski98} and was compared with the model predictions by \citet{Girardi01} for a metallicity of $z = 0.0025$. Investigating the stars with $E(V-I)_{RR} < 0.03$\,mag we find a very small difference of $0.006$\,mag between the RR\,Lyrae reddening values and the RC method. Thus, within the uncertainties, the RC and RR\,Lyrae reddenings agree very well for regions with low reddening.
\subsection{Comparison with other studies}
\label{comparision_with}
Previous studies of the reddening of the MCs concentrated mostly on the LMC. Below we compare our results for the LMC with the reddening maps inferred from OGLE\,II data by \citetalias{Subramaniam05} (\citetalias{Subramanian09} is not publicly available), with \citetalias{Pejcha09}, who used RR\,Lyraes from the OGLE\,III survey, with the cool stars reddening map of \citetalias{Zaritsky04}, and with CO and infrared maps from \citetalias{Dobashi08}. For the comparison with the SMC we will concentrate on the reddening maps of \citetalias{Zaritsky02} and the CO and infrared measurements by \citetalias{Dobashi09}. \\\noindent\hspace*{1em}
\subsubsection{\citet{Dobashi08, Dobashi09}}
\citetalias{Dobashi08} and \citetalias{Dobashi09} have used infrared data from 2MASS to derive reddening maps of both MCs. The absolute reddening values are not published, but the images of the maps can be visually compared to the results of our study. The qualitative consistency between the overall appearance of the maps derived from infrared data and our $E(V-I)$ maps is very good. All regions with high reddening values in the infrared are well reproduced by the maps of our work (Figure 5 of \citetalias{Dobashi08} and \citetalias{Dobashi09}, respectively). \\\noindent\hspace*{1em}
As expected the very active star-forming region in the wing of the SMC, with many H\,II regions, is the area with the highest reddening, having reddening values up to $E(V-I)_{RC} = 0.16$\,mag in our map. This region is very prominent in the infrared maps of \citet[their Fig.1]{Wilke03} as well. In the SMC bar the region in the southwest corner has the highest reddening values (up to $E(V-I)_{RC} = 0.12$\,mag in our work) in each of the inspected maps. \\\noindent\hspace*{1em}
In the LMC the 30\,Doradus region is the most prominent heavily reddened feature in each map. The location of the regions with high reddening is very similar. Additionally, the very low reddening of the LMC bar region found in our analysis is reproduced very well by the infrared maps of \citetalias{Dobashi08}. Only one area of the LMC is not reproduced in \citetalias{Dobashi08}. In the southwest corner of the observed OGLE\,III field a region with higher reddening is visible in the maps from our study, but not in the infrared data. Searching for extended gaseous features in this part of the sky we can only identify a higher concentration of HI clouds \citep[see][]{Kim98}. However, there is very good agreement of our data with the reddening maps from the 2MASS catalogue.
\subsubsection{\citet{Subramaniam05}}
\begin{figure}
\centering 
 \includegraphics[width=0.50\textwidth]{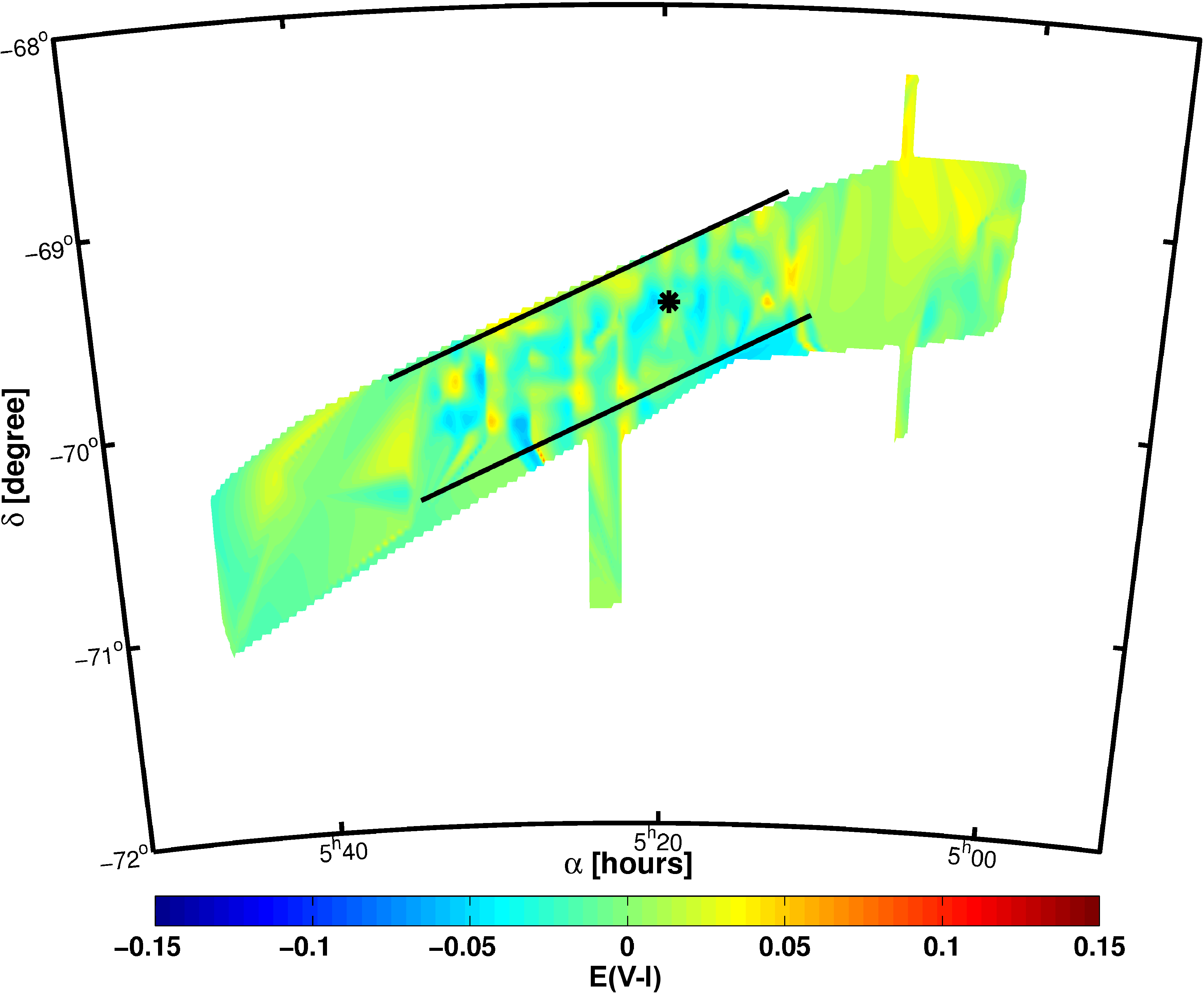}
 \caption{Difference map showing the reddening differences between \citetalias{Subramaniam05} and the red clump values in our study ($E(V-I)_{\mathrm{S05}} - E(V-I)_{\mathrm{RC/HGD10}}$. The region shown is the area covered by OGLE\,II in the LMC. Positive values show higher reddening inferred by \citetalias{Subramaniam05}. Negative values correspond to higher values in our dataset. Overall the differences are very small.} 
 \label{comparison_subramaniam} 
\end{figure}
The differences in the photometry between OGLE\,II and OGLE\,III are marginal (\citet{Udalski08a} quote $\sim 0.001$\,mag). Hence we do not expect that the use of OGLE\,II or OGLE\,III data within the same area will introduce differences. \\\noindent\hspace*{1em}
The comparison between \citetalias[their Figure\,1 and Figure\,17]{Subramaniam05}, from OGLE\,II data, and our study shows very good agreement (see Figure \ref{comparison_subramaniam}). The mean difference of the RC reddening values is $E(V-I) = 0.017$\,mag and therefore much smaller than the uncertainties. The fluctuations in the central parts are most likely caused by different subfield sizes. \citetalias{Subramaniam05}'s spatial resolution is a bit higher, hence the differences could be generated by variations on very small scales. For the northwestern area of the \citetalias{Subramaniam05} reddening map our RC reddening values are lower. While \citetalias{Subramaniam05} is limited by the OGLE\,II field our selection box includes many stars previously not observed. This difference in field selection might be the reason for the trend of slightly higher reddening in this region and in \citetalias{Subramaniam05} in general. Overall the differences are very small and our results are in very good agreement with the data of \citetalias{Subramaniam05}. \\\noindent\hspace*{1em}
\subsubsection{\citet{Pejcha09}}
Based on RR\,Lyrae stars from the OGLE\,III survey \citetalias{Pejcha09} use the period-colour relation of the unreddened cluster M3 to infer the $(V-I)_0$ of the LMC RR\,Lyraes. M3 has a metallicity of $[Fe/H] = -1.39$\,dex \citep{Cohen05}, similar to the mean metallicity of the LMC RR\,Lyrae \citep{Borissova09}. The difference between the theoretical and the measured colour yields the reddening. \citetalias{Pejcha09} subdivided the OGLE area and calculated a mean reddening for every field with at least two RR\,Lyrae stars. With this method \citetalias{Pejcha09} construct a map for nearly the whole LMC OGLE field of view. \\\noindent\hspace*{1em} 
\begin{figure}
\centering 
 \includegraphics[width=0.50\textwidth]{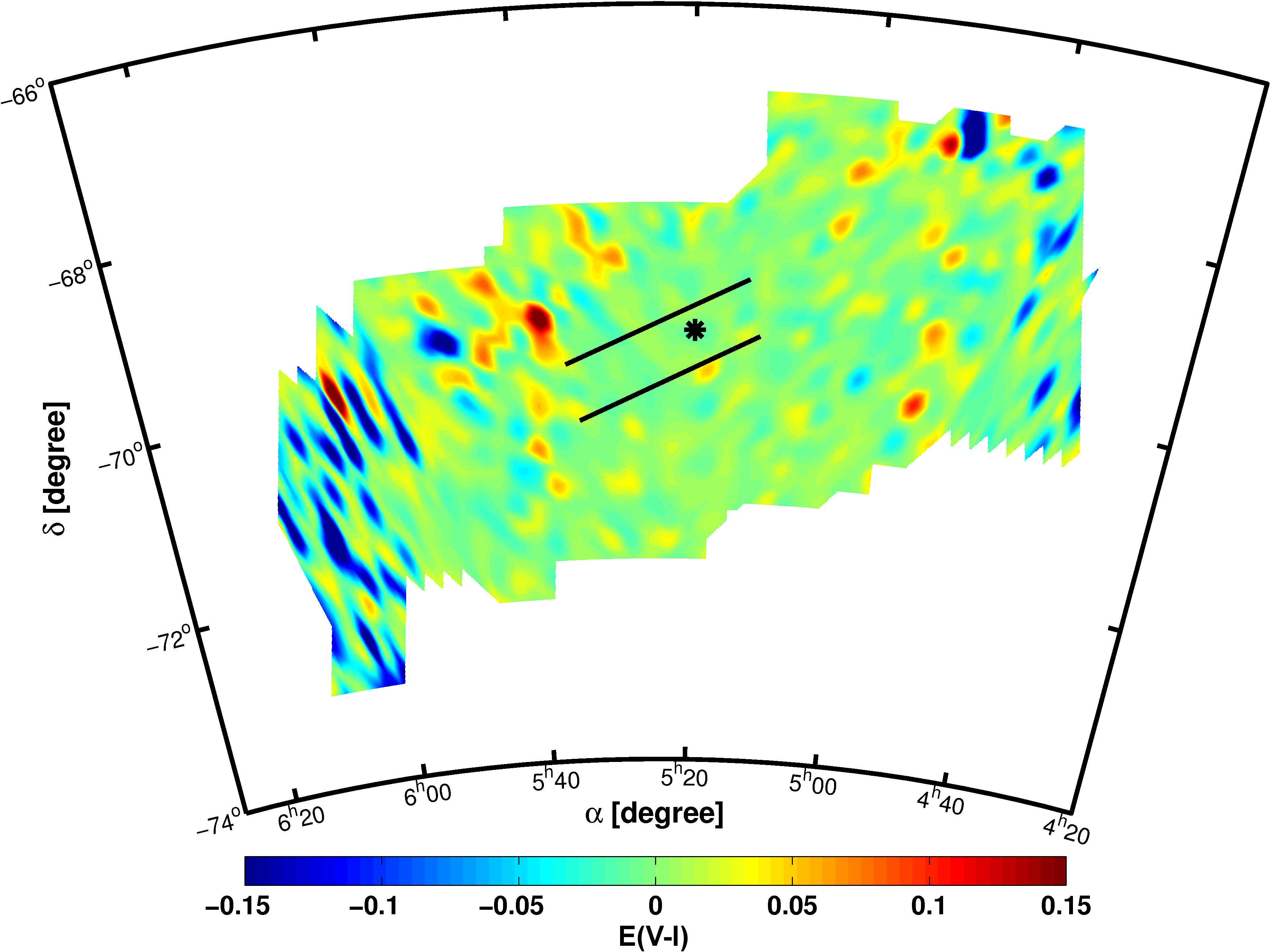}
 \caption{Difference map showing the reddening comparison between \citetalias{Pejcha09} and our study of the RR\,Lyrae stars. The plot shows $E(V-I)_{\mathrm{PS09}} - E(V-I)_{\mathrm{RR/HGD10}}$. Blue colours indicate higher reddening values in our calculations. We find a very good agreement between the two different methods with a mean difference of $0.007$\,mag. In 30\,Doradus the values by \citetalias{Pejcha09} are higher than in our estimates.} 
 \label{comparison_RR_pejcha} 
\end{figure}
While we used the same stars as \citetalias{Pejcha09}, there are two main differences in actually calculating the reddening values. First we accounted for the intrinsic metallicity of the RR\,Lyrae in our colour estimation. For different metallicity bins the mean colour varies significantly. The metallicity bin with the densest population and values between $-1.5 < [Fe/H] \leq -1.0$ has a mean colour of $(V-I) = 0.48$\,mag. All other metallicity bins have a redder colour (see Table\,\ref{Colour_FeH_table}). \\\noindent\hspace*{1em}
\begin{table}
\begin{center}
\caption{Mean colour of the RR\,Lyraes in certain metallicity bins.\label{Colour_FeH_table}}
\begin{tabular}{cc}
\tableline\tableline
metallicity bin & colour  \\    
\tableline                       
$-2.0 < [Fe/H] \leq -1.5$ & 0.52 \\
$-1.5 < [Fe/H] \leq -1.0$ & 0.48 \\ 
$-1.0 < [Fe/H] \leq -0.5$ & 0.51 \\ 
$-0.5 < [Fe/H] \leq -0.0$ & 0.54 \\ 
\tableline                                  
\end{tabular}
\end{center}
\end{table}
The other major difference is the formula used for calculating the colour of each RR\,Lyrae star. With a mean period $\mathrm{P} = 0.57$\,days and mean metallicity of $-1.23$\,dex for all RR\,Lyrae stars we obtain a mean colour of $0.49$\,mag. The relation $(V-I)_0 = 0.69 + 0.89\,\mathrm{P}$ (Equation\,1 in \citetalias{Pejcha09}) leads to a mean colour of $1.20$\,mag when using the same mean period. Compared with Figure\,\ref{red_clump_in_CMD} this would place the RR\,Lyrae stars redwards of the red clump. \\\noindent\hspace*{1em}
\citetalias{Pejcha09} transformed their reddening values to extinctions $\mathrm{A}_I$, which they made publicly available. We reverse the transformation and compare the reddening values $E(V-I)$ of \citetalias{Pejcha09} with our mean reddening values of the RR\,Lyrae that are observed in each field via $E(V-I)_{\mathrm{PS09}} - E(V-I)_{\mathrm{RR/HGD10}}$. \\\noindent\hspace*{1em}
On average we find that our reddening values agree very well with the reddening map presented by \citetalias{Pejcha09} (see Figure \ref{comparison_pejcha}). The mean difference is just $-0.004$\,mag, with deviations ranging from $-0.26$ to $0.22$\,mag. For $11\%$ of all \citetalias{Pejcha09} fields a deviation of more than 0.06\,mag or $1\sigma$ is found. The eastern regions of the LMC ($\alpha > 6^h$) are only sparsely populated and due to the selection criteria of \citetalias{Pejcha09} these regions are not very well covered. Therefore the comparison in these regions is not reliable. \\\noindent\hspace*{1em}
\begin{figure}
\centering 
 \includegraphics[width=0.50\textwidth]{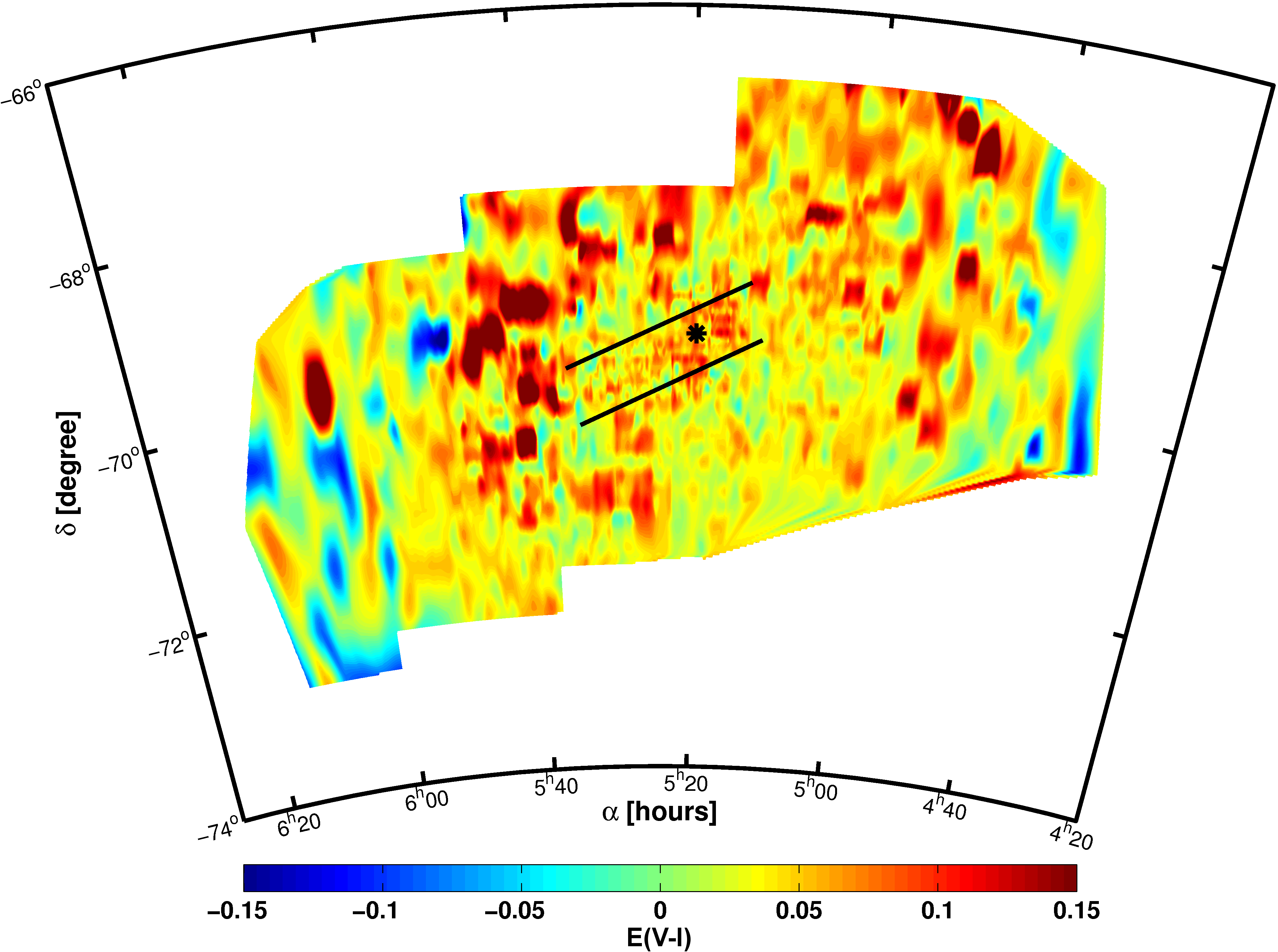}
 \caption{Difference map of the RR\,Lyrae-based reddening values of \citetalias{Pejcha09}, $E(V-I)_{\mathrm{PS09}}$, minus our reddening values derived from the RC, $E(V-I)_{\mathrm{RC/HGD10}}$. Overall the mean reddening of the map by \citetalias{Pejcha09} is enhanced by 0.05\,mag.} 
 \label{comparison_pejcha} 
\end{figure}
In addition to comparing the reddening values resulting from the different implementations of the RR\,Lyrae methods of \citetalias{Pejcha09} and our work, we compare the values of \citetalias{Pejcha09} to our reddening values from the RC method. We use the values of $E(V-I)$ computed above from the work of \citetalias{Pejcha09} and subtract our reddening estimates of the RC method for the corresponding field. The field sizes of the two approaches differ significantly. We therefore recalculate the values of \citetalias{Pejcha09} to fit into the spatial grid set by our RC method. The differences are evaluated using this grid and are shown in Figure\,\ref{comparison_pejcha}. This leads to a mean difference in reddening of $\overline{E(V-I)_{\mathrm{PS09}} - E(V-I)_{\mathrm{RC/HGD10}}} = 0.05$\,mag. The differences for individual fields range from $-0.17$ to $0.32$\,mag, but only $3\%$ of all fields have deviations of more than $0.12$\,mag or $2\sigma$. \\\noindent\hspace*{1em}
Especially the region of 30\,Doradus reveals much higher reddening values in the maps of \citetalias{Pejcha09}. Comparing the two RR\,Lyrae methods the same result is found, see Figure\,\ref{comparison_pejcha}. \\\noindent\hspace*{1em}
The northwestern region is higher in reddening as well in \citetalias{Pejcha09}. This result is also visible in the comparison of our RR\,Lyrae reddening estimates and the RC method (see Figure\,\ref{comparison_RR_RC_LMC}). \\\noindent\hspace*{1em}
Most regions with higher reddening values from the \citetalias{Pejcha09} map show indications of considerable differential reddening  \citepalias[compare with Figure\,2 in][]{Pejcha09}. As before we suggest that the most likely explanation is the differential reddening with its resultant asymmetric distribution of reddening values. Taking the mean of such a distribution yields higher values than the use of the peak of the distribution, as done for the RC stars. In addition, the sparser sampling of RR\,Lyrae stars as compared to the RC stars increases the scatter (see Section\,\ref{RC_colour_two}).
\subsubsection{\citet{Zaritsky02, Zaritsky04}}
\begin{figure}
\centering 
 \includegraphics[width=0.50\textwidth]{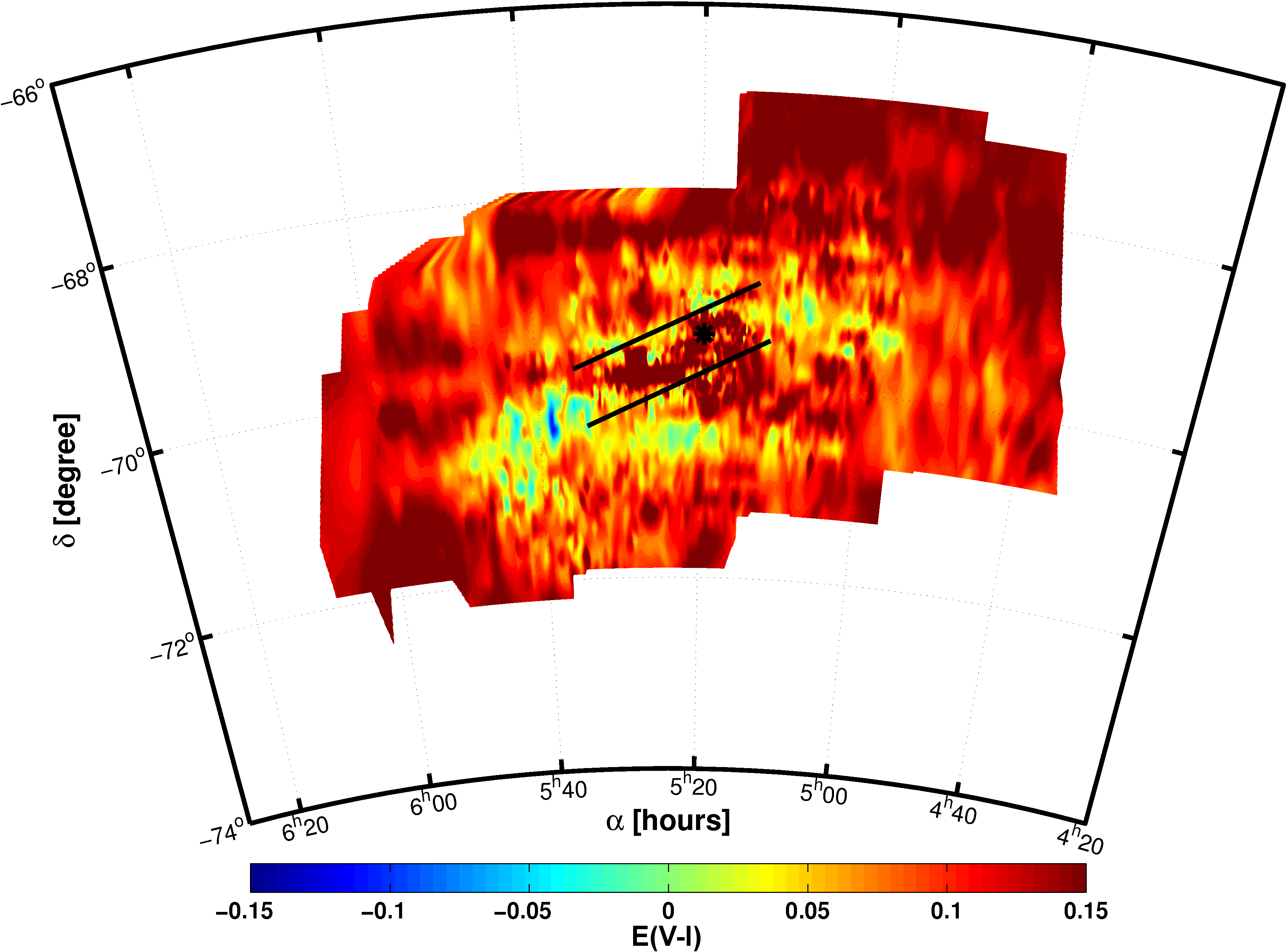}
 \caption{Difference map of the reddening values from the Magellanic Clouds Photometric Survey (\citetalias{Zaritsky04}) for cool stars are compared with our reddening values from the RC method, $E(V-I)_{\mathrm{Z04}} - E(V-I)_{\mathrm{RC/HGD10}}$. We averaged the \citetalias{Zaritsky04} reddening in the fields defined by the RC method. Positive values indicate higher reddening in the work of \citetalias{Zaritsky04}. The median difference of the two maps is 0.11\,mag.} 
 \label{comparison_Zaritsky_lmc} 
\end{figure}
\begin{figure}
\centering 
 \includegraphics[width=0.50\textwidth]{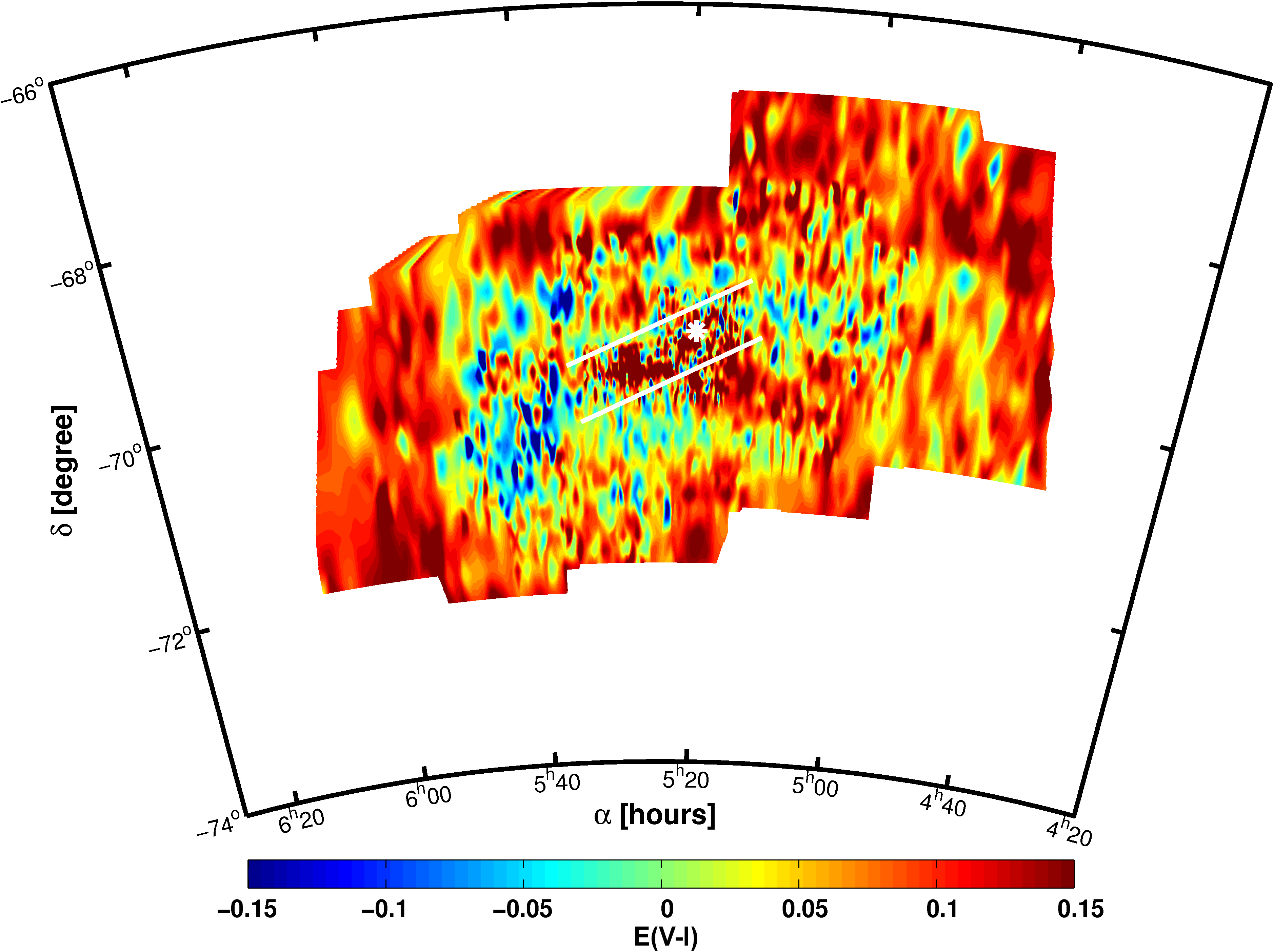}
 \caption{Difference map of the reddening determined by \citetalias{Zaritsky04}, $E(V-I)_{\mathrm{Z04}}$, for cool stars, minus our RR\,Lyrae based reddening, $E(V-I)_{\mathrm{RR/HGD10}}$. The resulting map shows a generally higher reddening of \citetalias{Zaritsky04} by 0.07\,mag. The reason for the high reddening residual in the centre of the LMC remains unclear. It has not be found in any other study.} 
 \label{comparison_RR_Zar} 
\end{figure}
\begin{figure}
\centering 
 \includegraphics[width=0.50\textwidth]{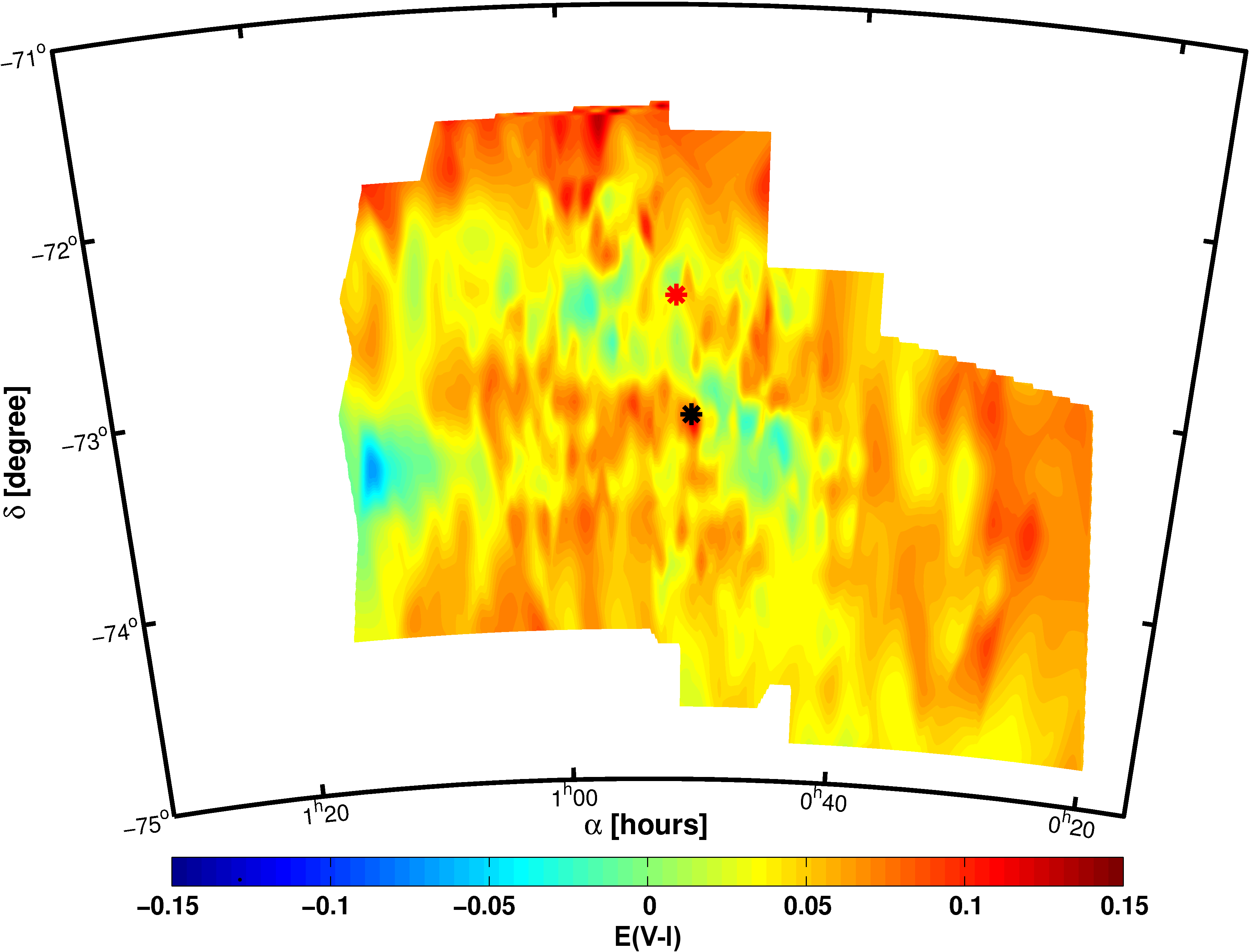}
 \caption{Difference map of our reddening values subtracted from \citetalias{Zaritsky02}'s reddening maps derived from the cool stars in the SMC. Despite the small offset to higher values in \citetalias{Zaritsky02}s data, which is still within our estimated errors, the agreement is very good. The mean difference is 0.04\,mag.} 
 \label{comparison_Zaritsky_smc} 
\end{figure}
\begin{figure}
\centering 
 \includegraphics[width=0.50\textwidth]{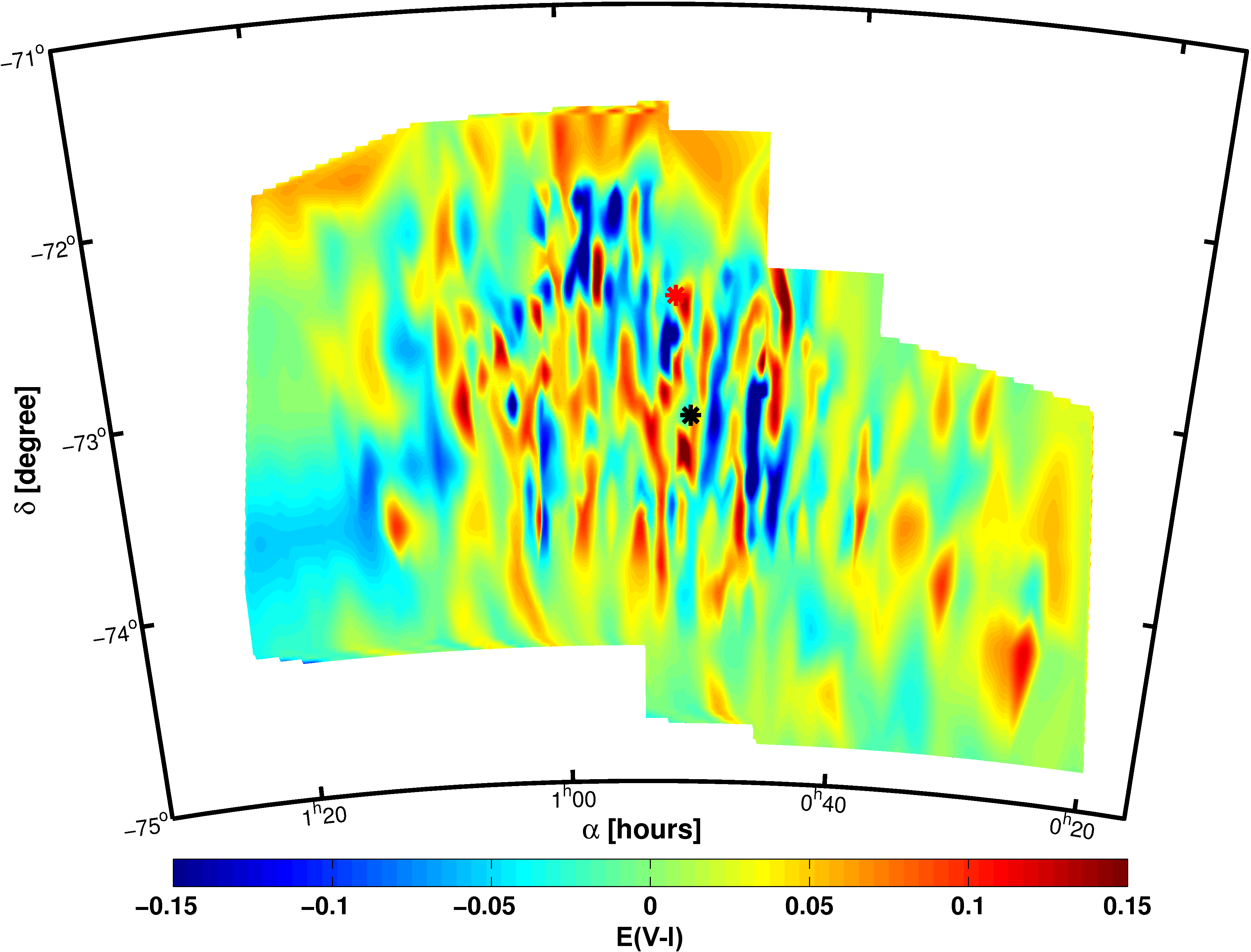}
 \caption{Difference map of the reddening determined by \citetalias{Zaritsky02}, $E(V-I)_{\mathrm{Z02}}$, for cool stars, minus our RR\,Lyrae based reddening, $E(V-I)_{\mathrm{RR/HGD10}}$. The resulting map shows very good agreement between the two different methods although the fluctuations are now larger than in Figure\,\ref{comparison_Zaritsky_smc}. The mean difference is basically zero.} 
 \label{comparison_RR_Zar_SMC} 
\end{figure}

\citetalias{Zaritsky02}(SMC) and \citetalias{Zaritsky04}(LMC) used hot ($12000K < T_E < 45000K$) and cool ($5500K < T_E < 6500K$) stars to produce reddening maps of both MCs. These maps are based on the multi-colour data of the MCPS \citep{Zaritsky97} and cover 64 square degrees in the LMC and 18 square degrees in the SMC \footnote{The fits files containing the reddening data of the MCPS can be found at: \url{http://ngala.as.arizona.edu/dennis/mcsurvey.html}}. For each hot and each cool star in the field a reddening value is estimated, but we only use the data of the cool stars, which are much closer to the characteristics of the RC,  with $T \sim 5000$\,K \citep{Bragaglia01a} and RR\,Lyrae stars, with $T \sim 6500$\,K \citep{Barcza10} at minimum light, used in our work. These reddening values are compared with all data sets evaluated by us.  \\\noindent\hspace*{1em}
For better comparability of the differences we compute the averages of \citetalias{Zaritsky04}'s reddening values in the 3174 LMC fields used for the RC method and subtract $E(V-I)_{\mathrm{RC/HGD10}}$ from the resulting $E(V-I)_{\mathrm{Z04/mean}}$. The comparison maps for the LMC are shown in Figure\,\ref{comparison_Zaritsky_lmc} and Figure\,\ref{comparison_RR_Zar}. The reddening maps for the LMC reveal differences between our two approaches and the \citetalias{Zaritsky04} estimate that exceed the errors found for the RC and RR\,Lyrae reddening values. While the central parts of the LMC are in general agreement, the reddening values for the less populated parts are considerably higher in the \citetalias{Zaritsky04} calculations. The origin of these differences remains unclear. Two major differences have to be taken into consideration: the different kinds of stars used and the different methods. \citetalias{Zaritsky02} and \citetalias{Zaritsky04} used stellar atmosphere models \citep[from][]{Lejeune97} to calculate the reddening for each star, while we use models of the location of RC stars in a CMD to assume absolute magnitudes. \\\noindent\hspace*{1em}
The mean difference between our RC map and \citetalias{Zaritsky04}'s values is 0.10\,mag. For only 2\% of all fields the mean reddening of the \citetalias{Zaritsky04} stars located therein is lower than the corresponding RC reddening. The reddening differences range from $-0.11$ to 0.46\,mag. About $37\%$ of all fields differ by more than 0.12\,mag. \\\noindent\hspace*{1em} 
In order to compare our RR\,Lyrae reddening values with \citetalias{Zaritsky04}'s reddening estimates for the LMC we use the equation $E(V-I)_{\mathrm{Z04}} - E(V-I)_{\mathrm{RR/HGD10}}$. As for the RC reddening we average the RR\,Lyrae and the \citetalias{Zaritsky04} reddening in the grid set up for the RC method. This allows us to compare not single star values but mean values with each other. \\\noindent\hspace*{1em}
The mean difference is found to be 0.07\,mag (Figure \ref{comparison_RR_Zar}), while the differences reach extremes of up to 0.50\,mag and down to $-0.40$\,mag. $17\%$ of all fields have negative values, while $21\%$ have discrepancies larger than $2\sigma$. As for the comparison with the RC method, these two maps are not in very good agreement. Especially the offset for the central region is striking. The origin of the rather high reddening residuals of \citetalias{Zaritsky04} in the centre of the LMC (Figure\,\ref{comparison_Zaritsky_lmc} and \ref{comparison_RR_Zar}) remains unclear. It has no counterpart in other reddening maps in the literature. \\\noindent\hspace*{1em}
In contrast the agreement in the reddening values of the SMC derived via the RC method and the \citetalias{Zaritsky02} values is very good. We compute $E(V-I)_{\mathrm{Z02}} - E(V-I)_{\mathrm{RC/HGD10}}$ and find that the median difference is only 0.04\,mag, while a general slight trend to higher values is obvious (Figure\,\ref{comparison_Zaritsky_smc}). But this offset is still well within the errors stated for the SMC reddening in our work. For the SMC about $2\%$ of the stars have lower reddening in \citetalias{Zaritsky02} than the corresponding value from the RC method and only 2 fields have differences larger than 0.12\,mag. The extremes range from $-0.07$\,mag to 0.13\,mag. \\\noindent\hspace*{1em}
Comparing the reddening maps from \citetalias{Zaritsky02} and the RR\,Lyrae we find even better agreement (Figure\,\ref{comparison_RR_Zar_SMC}) if we average the RR\,Lyrae values in the fields defined by the RC method. The difference $E(V-I)_{\mathrm{Z02}} - E(V-I)_{\mathrm{RR/HGD10}}$ is in agreement with zero, while the individual fields vary between $-0.28$\, and 0.18\,mag. The number of RR\,Lyrae stars per RC field is very low and the small number statistics lead to variations. Therefore we do not interpret the small scale variations as real features. Anyhow we find that there is an overall good agreement between the new reddening maps and the maps provided by \citetalias{Zaritsky02}. Thus the reason for the large reddening differences for the LMC remains unsolved. 
\subsection{Other comparisons}
\begin{figure}
\centering 
 \includegraphics[width=0.50\textwidth]{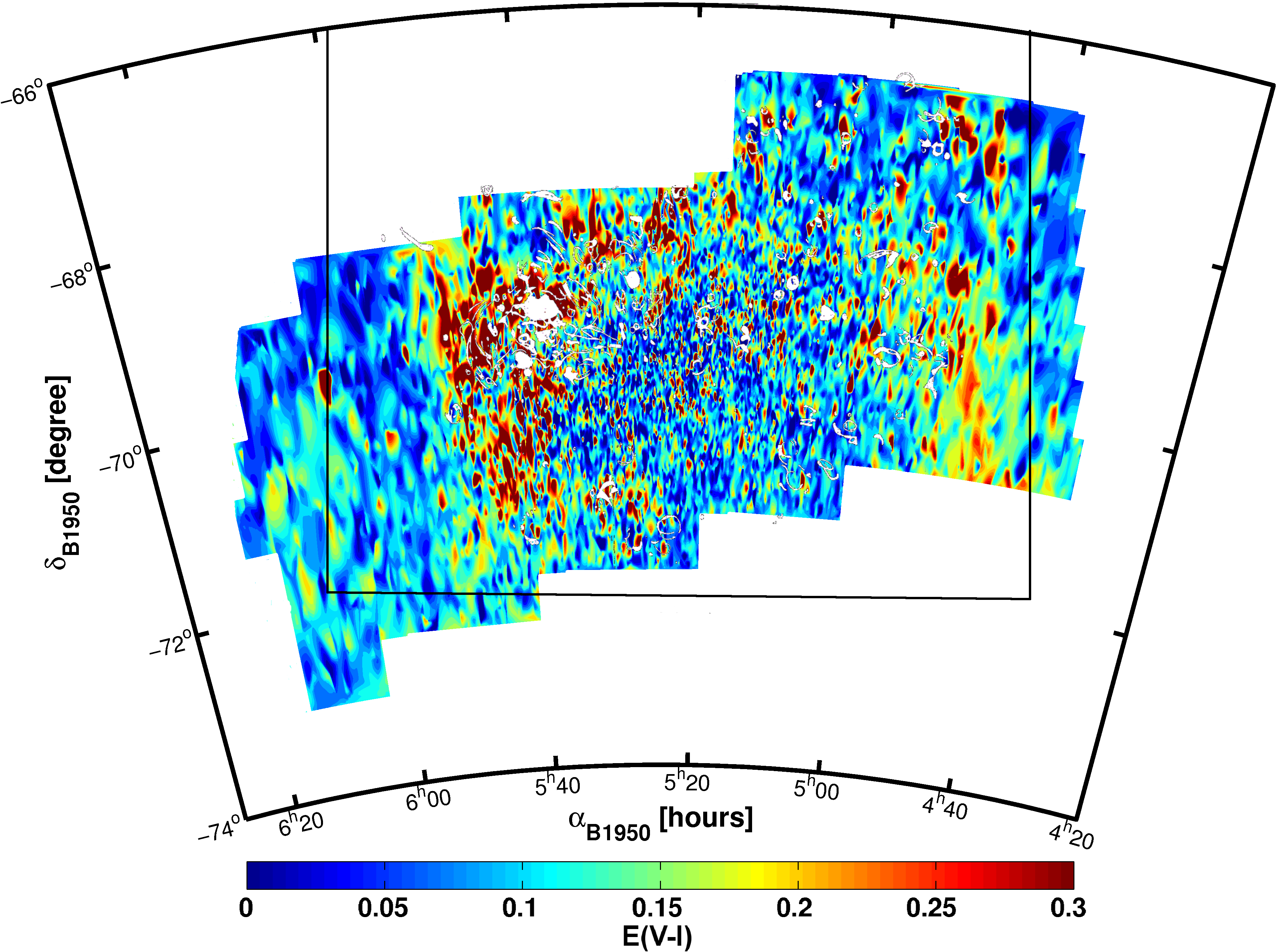}
 \caption{Superposition of the reddening map from Figure\,\ref{reddening_RR_LMC} and the map of H\,II regions by \citet[their Figure\,1]{Davies76}. The reddening enhancement is not conclusive by visual inspection, but the mean reddening at the centres of the H\,II regions is increased by 0.03\,mag.} 
 \label{comparison_RR_HII_lmc} 
\end{figure}
\begin{figure}
\centering 
 \includegraphics[width=0.5\textwidth]{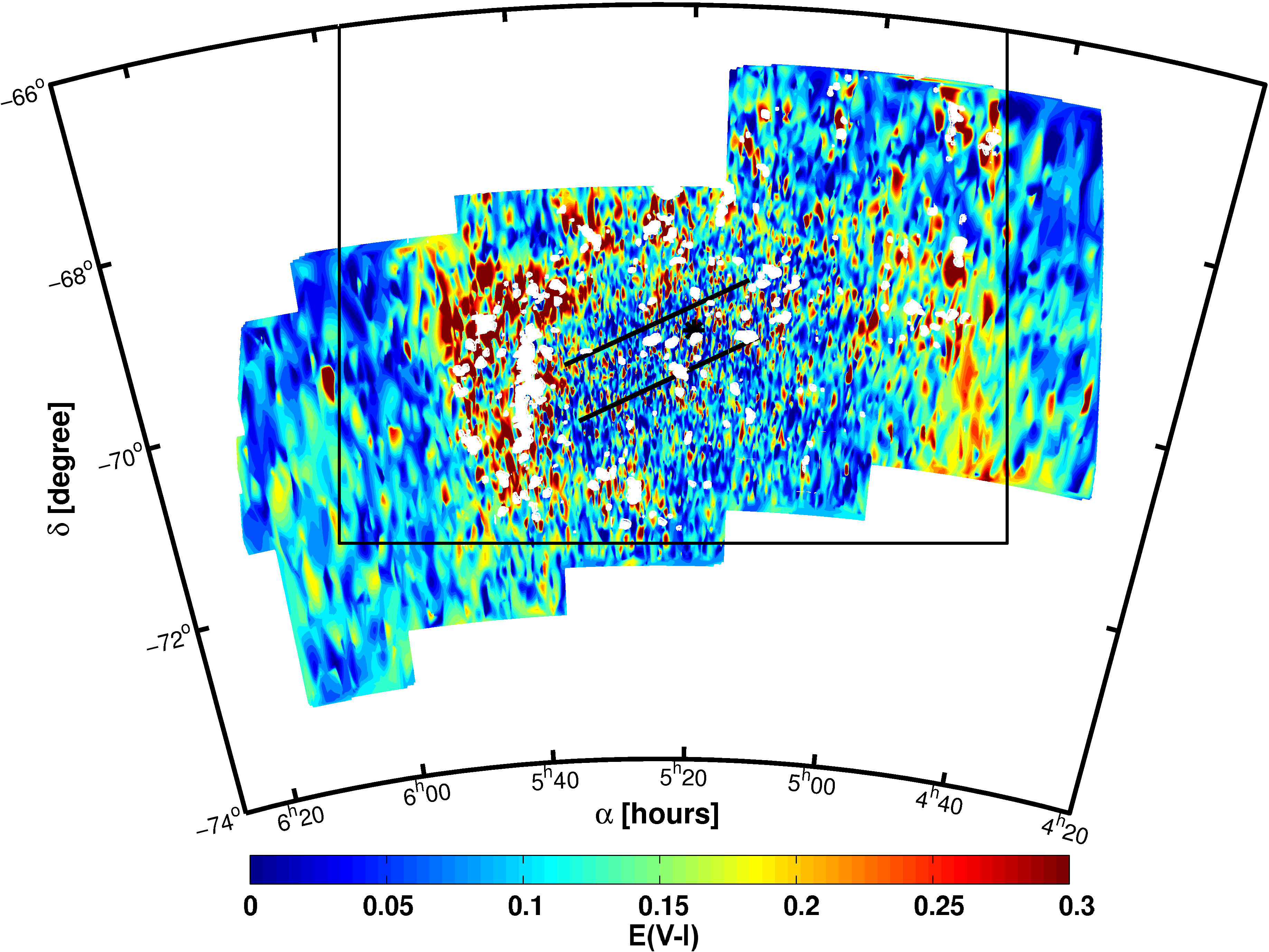}
 \caption{The reddening map from Figure\,\ref{reddening_RR_LMC} is superimposed with molecular clouds, traced by CO, from the catalogue by \citet[white contours]{Fukui08}. Overall we find an enhanced reddening at the spatial positions of the molecular clouds. The black lines represent the observed field of NANTES.} 
 \label{comparison_RR_CO_lmc} 
\end{figure}

In this section we will compare our reddening maps, mainly qualitatively, with the distribution of H\,II regions in the LMC \citep{Davies76, Book08, Book09} and SMC \citep{Davies76, Bica95}, with LMC CO maps, tracing molecular hydrogen clouds, by \citet{Fukui08} and the Spitzer dust maps by \citet{Meixner06} for the LMC and \citet{Bolatto07} for the SMC. \\\noindent\hspace*{1em}
We overlay the maps of \citet{Davies76} of the LMC and SMC with our reddening maps (Figure\,\ref{comparison_RR_HII_lmc}). Visual inspection does not always show a noticeable increase of reddening at the positions of the H\,II regions. We tentatively attribute this to effects of dust destruction by hot massive stars in the HII regions as well as to stellar winds that push gaseous and dusty material out of the forming cavity within the HII regions. Furthermore we compare the central positions of the H\,II regions, taken from \citet{Davies76, Bica95, Book08, Book09}, with the reddening values at that spatial location and find the reddening to be increased by 0.03\,mag for the LMC and SMC, respectively. \\\noindent\hspace*{1em}
In Figure\,\ref{comparison_RR_CO_lmc} the map of molecular hydrogen clouds as observed by NANTEN \citep{Fukui08}, and traced by CO, is superimposed with the reddening from the RR\,Lyrae stars. We find very good agreement of regions with enhanced reddening and the location of the clouds. This visual impression is verified by the computation of the mean reddening at the spatial positions of the molecular clouds. We find that the reddening is enhanced to a median value of $E(V-I)_{\mathrm{median}} = 0.19$\,mag. \\\noindent\hspace*{1em}
Finally we inspect the images of the SAGE (Surveying the Agents of Galaxy Evolution) survey of the LMC \citep{Meixner06} and find very good agreement of their $8\mu$m map with the LMC reddening map as obtained by us. The 8$\mu$m emission originates from small dust grains as traced by polycyclic aromatic hydrocarbons (PAHs) located in the outskirts of star-forming regions \citep{Meixner06}. Only the region of enhanced reddening in the south-west is not reproduced. However, we find good agreement of this region in the 160$\mu$m map of Spitzer. The absorbtion feature at larger wavelength are caused by greater particles. The 160 $\mu$m maps trace emission of relatively large and cool ($\sim 20$ K) dust grains at larger distances from star-forming regions \citep{Meixner06}. The SAGE-SMC \citep{Bolatto07} maps show very good agreement with the reddening map of the SMC as obtained by us.

%

\section{Summary}
\label{summary}

In this paper new optical reddening maps for the Magellanic Clouds are presented using two different methods based on OGLE\,III photometry. The first one uses the theoretical position of the RC in $\mathrm{(V-I)_0}$ colour in the CMD for a given mean metallicity. The observed shift of the measured RC position is used to determine the reddening. For this purpose we subdivide the original OGLE\,III fields into smaller fields and determine the mean colour of the RC in each of them. The size of the fields varies depending on the number of stars in each field. For the LMC we obtain 3174 subfields with a mean density of 1257 stars, while for the SMC the reddening is calculated in 693 subfields, with an average of 1318 stars per field.\\\noindent\hspace*{1em}
The maps show in general a low reddening. In the LMC we find a mean $E(V-I) = 0.09 \pm 0.07$\,mag and for the SMC $E(V-I) = 0.04 \pm  0.06$\,mag. Nevertheless some regions show a much higher reddening. The star-forming region 30\,Doradus in the LMC shows values up to $E(V-I) = 0.43$\,mag. The high differential reddening in this region broadens the distribution of the RC stars and there is an extended tail of even higher reddening values in this region. Another region with high reddening is found in the southwest of the examined OGLE\,III field and coincides with an H\,I region. In contrast to the highly extincted star-forming regions the reddening in the bar of the LMC is quite low.\\\noindent\hspace*{1em} 
For the SMC the RC method reveals three pronounced areas of higher reddening. The highest reddening is reached, with $E(V-I) = 0.16$\,mag, in an H\,II region in the wing. The other two regions are located in the bar. But overall the reddening is quite low.\\\noindent\hspace*{1em} 
For the second method the observed colour $(V-I)$ of RR\,Lyrae stars in the MCs is compared with the colour predictions of the calculated absolute magnitudes. These depend on the period and metallicity. The metallicity is evaluated via the Fourier decomposition of the lightcurve of each star. We obtain an independent reddening for each RR\,Lyrae star. For the LMC a median reddening of $E(V-I) = 0.11 \pm 0.06$\,mag is obtained. In the star-forming region 30\,Doradus the reddening reaches values up to $E(V-I) = 0.66$\,mag. The RR\,Lyrae reveal a mean reddening of the SMC of $E(V-I) = 0.07 \pm 0.06$\,mag.\\\noindent\hspace*{1em} 
The comparison with reddening maps derived from other studies show mostly good agreement. Moreover, a qualitative comparison with Spitzer maps shows areas of high reddening to coincide with regions of high dust emissivity from PAHs or larger dust grains at 8 and 160 $\mu$m, respectively. \\\noindent\hspace*{1em}
These new reddening maps show a consistent picture of the reddening distribution of both Magellanic Clouds. we make our position-dependent reddening values derived with the RC method  available via the German Astrophysical Virtual Observatory (GAVO) interface at \url{http://dc.zah.uni-heidelberg.de/mcx}.
%
%
\acknowledgments
We are thankful to the OGLE collaboration and to the MCPS collaboration for making their data publically available. We are obliged to Andreas Just for suggesting the RR\,Lyrae de-reddening. We are grateful to Markus Demleitner for helping us to incorporate the data into the virtual observatory and to the anonymous referee for the constructive comments to improve the manuscript.

\bibliography{Bibliography.bib}
\bibliographystyle{apj}

\end{document}